\documentclass[12pt,oneside]{report}

\usepackage{ai-sandbox-report}

\renewcommand{\ReportAuthors}{%
Muhammad Waseem, Md Aidul Islam, Md Nasir Uddin Shuvo,\\
Md Mahade Hasan, Kai-Kristian Kemell, Jussi Rasku, Mika Saari,\\
Vilma Saari, Roope Pajasmaa, Markku Oivo, and Pekka Abrahamsson
}

\renewcommand{\AuthorAffiliation}{%
GPT Lab, Faculty of Information Technology and Communication Sciences\\
Tampere University, Tampere, Finland\\
DIMECC Oy, Tampere, Finland\\
University of Oulu, Oulu, Finland
}

\renewcommand{\PdfTitle}{AI Sandbox: Architecture, Governance, and Prototype}
\renewcommand{\PdfAuthor}{Muhammad Waseem et al.}
\renewcommand{\PdfSubject}{Technical report on a secure and compliant AI sandbox}

\begin{document}



\providecommand{\ReportTitle}{AI Sandbox: Architecture, Governance, and Prototype}
\providecommand{\ReportType}{Technical Report}
\providecommand{\ReportDate}{March 1, 2026}
\providecommand{\VersionTag}{0.1}
\providecommand{\DocClass}{Public}

\providecommand{\ReportAuthors}{Muhammad Waseem, Md Aidul Islam, Md Nasir Uddin Shuvo, Md Mahade Hasan,
Kai-Kristian Kemell, Jussi Rasku, Mika Saari,
Vilma Saari, Roope Pajasmaa, Markku Oivo, Pekka Abrahamsson}

\providecommand{\AuthorAffiliation}{Faculty of Information Technology and Communication Sciences,
Tampere University, Tampere, Finland}

\providecolor{brand}{RGB}{25,55,109}
\providecolor{panel}{gray}{0.95}

\begin{titlepage}
\thispagestyle{empty}

\noindent\colorbox{brand}{%
  \parbox{\dimexpr\textwidth-2\fboxsep\relax}{%
    \vspace{14pt}
    \hspace{12pt}\color{white}\bfseries \ReportType
    \vspace{14pt}
  }%
}

\vspace{2.3cm}

\begin{center}
  {\Huge\bfseries \ReportTitle \par}
\end{center}

\vspace{2.2cm}

\begin{center}
  {\large\bfseries Authors}\par
  \vspace{8pt}
  {\large \ReportAuthors \par}
\end{center}

\vspace{1.2cm}

\begin{center}
  {\large\bfseries Affiliation}\par
  \vspace{6pt}
  {\large \AuthorAffiliation \par}
\end{center}

\vspace{1.2cm}

\begin{center}
  {\large\bfseries ISBN}\par
  \vspace{6pt}
  {\large \ReportISBN \par}
\end{center}

\vfill

\begin{center}
  {\small
  \ReportDate \quad | \quad Version \VersionTag \quad | \quad Classification: \DocClass
  }
\end{center}

\vspace{1.2cm}

\noindent\colorbox{panel}{%
  \parbox{\dimexpr\textwidth-2\fboxsep\relax}{%
    \vspace{8pt}
    \centering\small
    This document is provided for information purposes. Redistribution is permitted only according to the stated classification.
    \vspace{8pt}
  }%
}

\end{titlepage}
\clearpage

\pagenumbering{roman}
\chapter*{Acknowledgements}
\addcontentsline{toc}{chapter}{Acknowledgements}

The preparation of this report benefited from generative AI tools used for language polishing, consistency checks, formatting support, and the visualisation of a limited number of figures for improved readability.
The final manuscript was reviewed by the authors prior to release.

\clearpage
\chapter*{Revision History}
\addcontentsline{toc}{chapter}{Revision History}

\renewcommand{\arraystretch}{1.2}

\begin{tabularx}{\textwidth}{@{}p{1.3cm} p{2.5cm} p{2cm} p{1.8cm} X@{}}
\toprule
\textbf{Ver.} & \textbf{Date} & \textbf{Author(s)} & \textbf{Status} & \textbf{Summary of Changes} \\
\midrule

0.1 &
30 Sep 2025 &
MW, PA &
Draft &
Initial draft including system requirements, preliminary architecture definition, and baseline governance and policy framework for the GPT Lab Sandbox. \\

0.9 &
20 Jan 2026 &
MW, PA &
Internal Review &
Terminology consolidation; updates to governance and security sections; refinement of MVP definition and deployment documentation for CSC Rahti environment. \\

1.0 &
03 March 2026 &
MW, PA &
Release &
Final reviewed and approved release version. Architecture diagrams validated; references cross-checked; formatting and structural consistency finalized for publication. \\

\bottomrule
\end{tabularx}

\vspace{1em}

\noindent\textit{Note: All changes are documented in accordance with the internal quality assurance procedures of GPT Lab.}

\clearpage
\chapter*{Executive Summary}
\addcontentsline{toc}{chapter}{Executive Summary}

\setlength{\fboxsep}{4pt}

\fcolorbox{brand}{panel}{%
  \begin{minipage}{\dimexpr\textwidth-2\fboxsep-2\fboxrule\relax}

  \begingroup
  \setlength{\parindent}{0pt}
  \setlength{\parskip}{4pt}     
  \small                        

  Artificial intelligence development is increasingly constrained by regulatory, security,
  and data-governance requirements. Organisations must innovate rapidly while demonstrating
  compliance with the \textbf{GDPR}, the \textbf{EU AI Act}, and emerging cybersecurity
  obligations. However, trusted environments that combine experimentation speed with
  governance-by-design remain limited.

  This report presents the design and implementation of the \textbf{AI Sandbox for
  Industry--Academia Collaboration}: a secure, transparent, and compliance-ready platform
  enabling controlled AI experimentation across organisational boundaries. The Sandbox
  integrates technical architecture, governance mechanisms, and operational policies into a
  unified and auditable environment.

  \textbf{Scope of Work.} The project comprised:
  \begin{itemize}
    \setlength\itemsep{2pt}
    \setlength\topsep{4pt}
    \item Structured stakeholder-driven requirements analysis;
    \item Design of a modular, container-based system architecture;
    \item Development of a governance and policy framework aligned with European regulation;
    \item Implementation and deployment of a functional MVP on the CSC Rahti cloud platform.
  \end{itemize}

  \textbf{Key Outcomes.}
  \begin{itemize}
    \setlength\itemsep{2pt}
    \setlength\topsep{4pt}
    \item \textbf{Validated Requirements Framework:} Prioritised functional and non-functional requirements addressing data sensitivity, localisation constraints, security, and regulatory compliance.
    \item \textbf{Modular Architecture:} A scalable, interoperable, and container-based design comprising orchestration, AI service layers, data management, and protection mechanisms.
    \item \textbf{Governance-by-Design Model:} Clearly defined organisational roles, access control policies, audit mechanisms, and compliance procedures aligned with GDPR and the EU AI Act.
    \item \textbf{Operational MVP Deployment:} A working prototype demonstrating secure access control and managed experimentation workflows.
  \end{itemize}

  The Sandbox establishes a standards-aligned infrastructure where governance, compliance,
  and technical capability are integrated by design rather than applied retrospectively.

  \endgroup

  \end{minipage}
}
\clearpage

\tableofcontents\clearpage
\listoffigures\clearpage
\listoftables\clearpage

\pagenumbering{arabic}
\chapter{Introduction}

\fcolorbox{brand}{panel}{%
  \begin{minipage}{0.97\textwidth}
  \vspace{0.6em}
  \textbf{Highlights}
  \begin{itemize}
    \item Purpose of the Sandbox and this document
    \item Motivation and importance of secure industry--academia collaboration
    \item Prototype overview and demonstration of feasibility
    \item Thematic service areas of the Sandbox
    \item Stakeholders, onboarding processes, and collaboration flows
    \item Built-in compliance, monitoring, and security mechanisms
    \item Positioning of the Sandbox as a Minimum Viable Product (MVP)
  \end{itemize}
  \vspace{0.4em}
  \end{minipage}
}

\vspace{1.5em}

\section{Purpose of the Document}
This document introduces the AI Sandbox, a secure yet shared environment where companies, universities, research institutes, and public organisations can collaborate on research and development. The current version has been demonstrated as a \textbf{prototype}, showing the feasibility and value of such an environment.

The purpose of this report is to document the work completed so far, present the results from the prototype, and establish a foundation for the next phase of development. While the prototype illustrates the main features and structure, the system will evolve based on feedback, new requirements, and future investments.

\section{Background and Motivation}
There is a growing need for a trusted environment where industry and academia can safely experiment with AI, machine learning, and software engineering tools \cite{OECD2019AI,ENISA2020AIThreatLandscape}. Many organisations face difficulties in testing AI-driven solutions due to security, compliance, or cost constraints. Similarly, universities require practical ways to transfer research outputs into industrial practice.

The AI Sandbox addresses these needs by combining a secure technical setup with clear governance and thematic services. It lowers barriers for experimentation and collaboration while ensuring compliance with Finnish regulations, the EU AI Act, and GDPR \cite{EUAIAct2024,gdpr,FinDPA2018,NIS2}.

\section{Importance of Collaboration}
The Sandbox is built on the principle that innovation cannot happen in isolation \cite{Etzkowitz2000TripleHelix,Chesbrough2003OpenInnovation}. Companies, universities, research institutes, and public organisations each hold expertise needed to develop trustworthy and effective solutions.

\begin{itemize}
  \item \textbf{Companies} contribute domain knowledge, practical problems, and pathways to scale.
  \item \textbf{Universities and research institutes} provide cutting-edge methods, critical evaluation, and training of new talent.
  \item \textbf{Public organisations} ensure solutions align with societal needs, regulation, and public trust.
\end{itemize}

By creating a shared and secure space for these groups, the Sandbox maximises their strengths. It accelerates innovation, reduces risks by providing a controlled environment, and ensures outcomes meet both industrial and societal expectations.

\section{Stakeholder Consultations}
The design has been informed by direct input from:
\begin{itemize}
  \item \textbf{Three external company representatives}, including SMEs and industrial stakeholders.
  \item The \textbf{GPT-Lab research team} at Tampere University.
  \item \textbf{Bi-weekly meetings} with consortium representatives.
\end{itemize}

Their feedback was consolidated into six thematic areas that form the logical backbone of the Sandbox services (see Table~\ref{tab:themes}).

\begin{table}[H]
\centering
\scriptsize
\rowcolors{2}{white}{panel}
\begin{tabularx}{\textwidth}{p{3cm} p{5.5cm} X}
\toprule
\rowcolor{lightgray}
\textbf{Theme} & \textbf{Focus} & \textbf{Example Features} \\
\midrule

Research \& Development &
Controlled experimentation, reproducibility, benchmarking \cite{Sculley2015HiddenDebt,Breck2017MLTestScore,MLflow2018} &
Results dashboards, curated datasets, experiment tracking \\

AI \& Machine Learning &
Model training, automation, evaluation &
Text analysis, computer vision, speech processing, fine-tuning workflows \\

Templates \& Workflows &
Ready-made processes for industry and academia &
Project templates, domain-specific workflows, automated pipelines \\

Security \& Compliance \cite{ISO42001,ISO23894,NIST_AIRMf} &
Trust, neutrality, regulatory adherence &
GDPR dashboards, EU AI Act readiness, vulnerability scanning \\

Infrastructure \& Resources &
Scalable technical foundation &
GPU management, orchestration, secure data storage, automated testing \\

Collaboration \& Management &
Organisational and human collaboration &
Project tracking, resource sharing, collaboration analytics, scheduling \\

\bottomrule
\end{tabularx}
\caption[Thematic areas of the Sandbox]{Thematic areas of the Sandbox, derived from consultations.}
\label{tab:themes}
\end{table}

\section{System Prototype Overview}
The prototype demonstrates the Sandbox as a multi-tenant platform with a rich set of administrative and user-facing capabilities. Core features include:

\begin{itemize}
  \item A \textbf{System Admin Console} for global platform administration, with live health monitoring of CPU, memory, storage, and network usage, as well as real-time activity dashboards for users and services.
  \item \textbf{User management} tools for adding, approving, and monitoring roles such as researchers, students, and administrators, ensuring secure collaboration with clear responsibilities \cite{Sandhu1996RBAC}.
  \item \textbf{Organisation management} functionality, where universities, research institutes, and companies are registered as tenants, each with its own members and project environments.
  \item A \textbf{Service management interface} that enables administrators to deploy, configure, and monitor services such as the AI Services Platform, Data Catalog, Collaboration Hub, Analytics Dashboard, Security \& Compliance service, and Model Training Service.
  \item Multi-tenant awareness, shown in the distribution of roles (researchers, students, administrators) and services (AI, data, analytics, security, collaboration).
\end{itemize}

Together, these features illustrate a working backbone for experimentation and collaboration that integrates technical control, compliance, and scalability.

\section{Collaboration and Onboarding}
Collaboration and onboarding are central to the Sandbox vision:
\begin{itemize}
  \item \textbf{Company onboarding:} Organisations can register profiles, contribute resources, and access compliance-ready infrastructure.
  \item \textbf{Academic onboarding:} Researchers join via the Academic Hub, use project templates, and share results in controlled formats.
  \item \textbf{Collaboration matching:} Smart search and AI-powered recommendations enable discovery of partners and launch of projects.
  \item \textbf{Shared experimentation:} Teams jointly access secure execution environments while safeguarding data and compliance.
\end{itemize}

\section{Compliance and Security}
Even as a prototype, compliance and data protection have been built in:
\begin{itemize}
  \item Multi-tenant isolation at organisation, project, and service levels.
  \item Role-based access control and user approval workflows.
  \item Audit logging of administrative actions and service usage.
  \item Security \& compliance services with GDPR and EU AI Act monitoring.
\end{itemize}

These features ensure that while the environment is shared, each tenant’s data and experiments remain secure.

\section{Positioning and Value}
The AI Sandbox, in its prototype form, functions as a \textbf{Minimum Viable Product (MVP)}. It demonstrates end-to-end workflows: from organisation onboarding to user and service management, experimentation, and compliance monitoring.

Looking ahead, it will evolve into a fully operational environment that supports larger-scale pilots, long-term collaborations, and Nordic--Baltic initiatives. The prototype has provided the necessary proof to take these next steps and invest in a secure, collaborative, and future-ready innovation platform.

\chapter{Requirements and Data Analysis}

\fcolorbox{brand}{panel}{%
  \begin{minipage}{0.97\textwidth}
  \vspace{0.6em}
  \textbf{Highlights}
  \begin{itemize}
    \item Requirements elicited through structured consultations with Bittium, Q4US, Solita, DIMECC, and Tampere University GPT-Lab
    \item Prioritised use cases capturing real-world value for SMEs, industry, and academia
    \item Functional requirements covering AI services, user management, data handling, collaboration, and compliance
    \item Non-functional requirements addressing security, scalability, usability, neutrality, and legal compliance
    \item Domain-specific needs for healthcare, public services, and manufacturing
    \item AI Service Catalogue of 41 services grouped into five categories
    \item Requirement prioritisation and traceability to guide MVP, advanced, and strategic phases
  \end{itemize}
  \vspace{0.4em}
  \end{minipage}
}
\vspace{1.5em}

\section{Consultation Process and Stakeholders}

The requirements for the AI Sandbox were identified through a systematic and iterative
consultation process that combined interviews, workshops, prototype reviews, and sprint
demonstrations \cite{ISO29148}. The process was designed to capture both short-term needs related to the
Minimum Viable Product (MVP) and longer-term requirements for scaling the platform into a
more mature, enterprise-ready environment.

\subsection*{Consultation Approach}
\begin{itemize}
  \item \textbf{Structured Interviews:} One-on-one sessions of 30--45 minutes were
        conducted with company representatives and research partners to collect detailed
        functional and non-functional requirements.
  \item \textbf{Sprint Reviews:} At the end of development sprints, prototype features
        were demonstrated and feedback was gathered from stakeholders to refine
        requirements.
  \item \textbf{Bi-weekly Consortium Meetings:} Regular meetings ensured that academic,
        industrial, and coordination partners were aligned and that requirements evolved
        in response to feedback.
  \item \textbf{Evidence Collection:} Requirements were traced against artefacts,
        including prototype snapshots (e.g., admin console, user/organisation management)
        and system feature documentation \cite{ISO29148}.
\end{itemize}

\subsection*{Participating Organisations}
The following organisations contributed actively to the requirements gathering process:
\begin{itemize}
  \item \textbf{Bittium}:  A Finnish technology company specialising in secure
        communications and mission-critical software systems. Bittium emphasised the
        importance of tenant isolation, compliance with security standards, and reliable
        benchmarking frameworks for AI models.
  \item \textbf{Q4US}:  A software engineering SME with a focus on applied AI solutions
        and early adoption of new technologies. Q4US committed to testing the sandbox MVP
        and stressed the need for cost estimation tools, benchmarking templates, and
        automated testing support.
  \item \textbf{Solita}:  A digital transformation company with expertise in
        data engineering, cloud platforms, and AI-driven services. Solita highlighted the
        importance of data governance, anonymisation, and the ability to integrate
        domain-specific datasets in compliance with country-specific laws.
  \item \textbf{DIMECC}:  Serving as the facilitator of the consortium, DIMECC provided
        coordination, ensured neutrality, and represented broader industrial engagement.
        Their input focused on governance, sustainability, and adoption models.
  \item \textbf{Tampere University GPT-Lab}:  The academic research partner responsible
        for LLM evaluation, architecture design, and alignment with EU AI Act and related
        standards. The GPT-Lab highlighted the importance of traceability, compliance
        monitoring, and reproducible experimentation.
\end{itemize}

\subsection*{Key Findings}
The consultations revealed several recurring requirements that cut across different
stakeholder groups:
\begin{itemize}
  \item A secure sandbox environment that enables experimentation without risking
        production systems.
  \item Benchmarking and estimation tools to help SMEs evaluate competitiveness and
        predict project costs.
  \item GDPR-compliant data handling, including anonymisation and data residency controls, to
        facilitate safe use of sensitive datasets \cite{gdpr}.
  \item A multi-tenant, collaborative platform that allows companies, universities, and public organisations to run projects together while maintaining data isolation \cite{NIST80053r5,KubernetesMultiTenancy}.
  \item Support for alignment with the EU AI Act, including risk assessment and ongoing monitoring \cite{EUAIAct2024}.
\end{itemize}

\section{Data and Requirements}
This section outlines the core requirements and data foundations of the AI Sandbox.
It describes how open and domain-specific datasets are accessed, cleaned, and used for
AI experimentation under GDPR and EU AI Act considerations. It then presents the onboarding
and collaboration workflows, system and administration features, AI services, and
hardware management capabilities. Together, these elements link stakeholder needs to
datasets, workflows, and platform services, providing a baseline for future
expansion.

\subsection{Data management}

We have collected a range of open and domain-specific data sources that can be used within
the AI Sandbox to support research, testing, and fine-tuning activities. These sources include national portals such as Avoindata.fi, Statistics Finland, and THL’s Sotkanet database, as well as European-level resources such as the EU Open Data Portal \cite{AvoindataFI,StatFinPxWeb,THLSotkanetAPI,DataEuropa}. In addition, specialised archives like Findata and the Finnish Social Science Data Archive (FSD) provide controlled access to sensitive healthcare and social datasets, enabling experimentation under strict compliance safeguards \cite{Findata,FSD}. The purpose of using these datasets is to provide researchers, companies, and public organisations with representative and compliant data that supports use cases ranging from benchmarking and domain adaptation to compliance testing and service innovation. Within the sandbox environment, data can be accessed through APIs or direct downloads, preprocessed via pipelines for cleaning and anonymisation, and then integrated into AI workflows for training, fine-tuning, or evaluation. A consolidated summary of the identified data sources and their relevance to sandbox use cases is provided in the following table.

\begin{table}[H]
\centering
\scriptsize
\caption{Summary of Data Sources}
\begin{tabular}{p{3.2cm} p{3.0cm} p{5.0cm} p{3.0cm}}
\toprule
\textbf{Source} & \textbf{Domain} & \textbf{Purpose / Use Case} & \textbf{Accessibility} \\
\midrule
Avoindata.fi (Finland’s Open Data Portal) & Public services, municipalities & Provides open datasets on government services, municipal operations, infrastructure, and demographics for AI-based public service innovation. & Open access via API and downloads \\
Statistics Finland (Tilastokeskus) & Demographics, economy, society & Statistical datasets on population, labour, economy, and social indicators for benchmarking and model evaluation. & Open access, bulk downloads \\
THL Sotkanet Database (Finnish Institute for Health and Welfare) & Healthcare, welfare & National indicators on healthcare, welfare, and wellbeing, enabling health-related AI experiments under compliance safeguards. & Open API access \\
Findata (Health and Social Data Permit Authority) & Sensitive healthcare and social data & Controlled access to individual-level register data for advanced health research and domain-specific fine-tuning. & Permit-based, regulated access \\
Finnish Social Science Data Archive (FSD) & Surveys, social research & Provides curated datasets from surveys and studies, supporting domain-specific experimentation in public and social services. & Open or restricted depending on dataset \\
EU Open Data Portal (data.europa.eu) & Cross-domain, EU-wide & Aggregates European datasets across governance, economy, environment, and policy for multilingual and cross-border AI use cases. & Open access via API and downloads \\
ELIXIR Finland / CSC Resources ELIXIR \cite{ELIXIRFinland,CSCElixir} & Life sciences, biomedical research & Offers domain-specific datasets and bioinformatics resources for healthcare-related AI experiments and model validation. & Access via research collaboration agreements \\
\bottomrule
\end{tabular}
\label{tab:datasources}
\end{table}

\subsection{Onboarding, Access \& Collaboration Workflows}

The AI Sandbox implements structured workflows that ensure smooth entry for
universities, companies, and individual researchers, while also supporting their
collaborative activities once onboarded. Each workflow provides tailored access
paths, compliance enforcement, and role-based resource allocation, ensuring that
participants can join and collaborate securely and efficiently.

\begin{enumerate}

  \item \textbf{Universities \& Research Institutes Onboarding}: Provides a dedicated academic hub where universities, faculties, and research groups can manage projects, collaborate with students, and handle research data. Includes publication support and structured onboarding paths for different types of academic users. Ensures that research activities are compliant, secure, and well-documented.

  \item \textbf{Companies \& Organisations Onboarding}: Enables corporate registration with support for R\&D collaboration, innovation projects, and technology transfer. Organisations can set up profiles, manage resources, and coordinate projects with academic partners. Compliance and governance features ensure corporate participation meets regulatory and industry requirements.

  \item \textbf{Individual Researchers Onboarding}: Provides independent researchers, consultants, and visiting scholars with a streamlined onboarding path. Enables participation in research projects, applications for roles, and professional networking opportunities. Ensures individuals can contribute meaningfully without requiring institutional backing.

  \item \textbf{Collaboration Categories}: Once onboarded, users can collaborate through structured categories: Company to Company, University to University, Company to University, Individual to Company, Individual to University, and Individual to Individual/Others. Each category provides workflows for project creation, partnerships, and knowledge exchange, enabling flexible yet governed collaborations across stakeholder types.

  \item \textbf{Self-Service Registration}: Allows users to register directly via the landing page with email verification, role selection, and organisation assignment. Reduces entry barriers while maintaining structured access for different user types. Approval workflows adapt to role sensitivity to ensure proper governance.

  \item \textbf{Role-Specific Entry Points}: Guides each user type through tailored onboarding routes (academic, corporate, or independent). Aligns the onboarding process with role expectations to prevent misuse and provide clarity. Automates compliance and resource allocation based on user type.

  \item \textbf{Free Trial Access}: Offers a 14-day trial period without financial commitments, lowering barriers for new users. Builds trust with prospective organisations by allowing them to explore the platform risk-free. Encourages experimentation and adoption at an early stage.

  \item \textbf{Compliance-Aware Access}: Embeds GDPR, EU AI Act, and Finnish regulatory requirements directly into the onboarding process. Classifies users according to compliance obligations and tailors access accordingly \cite{gdpr,EUAIAct2024}. Ensures legal requirements are addressed from the start of the user journey.

  \item \textbf{Approval Workflow Integration}: Balances accessibility with governance by automating low-risk approvals while requiring manual checks for sensitive roles. Supports multi-level organisational approval chains with escalation where needed. Provides oversight for high-privilege accounts and sensitive operations.

  \item \textbf{Authentication Controls}: Employs JWT-based authentication, strong password policies, and multi-factor authentication for secure login. Session management ensures identities are safeguarded throughout usage. Forms the foundation of secure access for all users \cite{RFC7519,NIST80063B4}.

  \item \textbf{Personalised Dashboards}: Directs users to dashboards tailored to their role, subscription tier, and resource allocation once onboarding is complete. Provides immediate access to relevant tools and data, reducing friction during adoption. Enhances productivity from the first login by contextualising the user experience.
\end{enumerate}

\subsection*{Analysis of Onboarding, Access \& Collaboration Workflows}

The following table provides a consolidated analysis of the onboarding and access
features in the AI Sandbox. These workflows define how different types of users
(universities, companies, and individuals) join the platform, gain access, and
collaborate with others. Each feature has been evaluated in terms of its value,
expected outcome, target domain, and implementation priority.

\begin{table}[H]
\centering
\scriptsize
\caption[Onboarding, Access]{Onboarding, Access \& Collaboration Workflows in the AI Sandbox}
\begin{tabular}{p{0.8cm} p{4.0cm} p{5.0cm} p{3.0cm} p{2.0cm}}
\toprule
\textbf{\#} & \textbf{Workflow Feature} & \textbf{Value / Outcome} & \textbf{Domain} & \textbf{Priority} \\
\midrule
1 & Universities \& Research Institutes Onboarding & Dedicated academic hub with project management, student collaboration tools, research data handling, and publication support & Academia, research institutes & High \\
2 & Companies \& Organisations Onboarding & Corporate registration with support for R\&D projects, technology transfer, and innovation partnerships & Industry, R\&D teams & High \\
3 & Individual Researchers Onboarding & Streamlined onboarding for consultants, postdocs, and independent researchers with project participation and networking tools & Freelance, academic visitors & Medium \\
4 & Collaboration Categories & Structured models for Company ↔ Company, University ↔ University, Company ↔ University, and Individual ↔ others, enabling flexible partnerships & Cross-domain collaboration & High \\
5 & Self-Service Registration & Direct registration with email verification, organisation assignment, and role selection to reduce entry barriers & All users & High \\
6 & Role-Specific Entry Points & Tailored onboarding routes (academic, corporate, individual) ensuring clarity, compliance, and resource alignment & All domains & High \\
7 & Free Trial Access & 14-day trial period to lower barriers for adoption and build trust with organisations before commitment & All prospective users & Medium \\
8 & Compliance-Aware Access & GDPR/EU AI Act aligned onboarding with user classification, consent, and data residency enforcement & Academia, industry, public sector & High \\
9 & Approval Workflow Integration & Automated low-risk approvals with manual checks for sensitive roles and multi-level organisational chains & All governance levels & High \\
10 & Authentication Controls & Secure login with JWT, password policies, MFA, and session management to safeguard identities & All domains & High \\
11 & Personalised Dashboards & Role-based dashboards ensuring immediate access to relevant tools, projects, and data after onboarding & All users & High \\
\bottomrule
\end{tabular}
\label{tab:onboarding}
\end{table}

\subsection{System \& Administration Features}

The AI Sandbox includes a comprehensive set of system and administration features
that support secure and compliant management of users, organisations, and
resources. These features form the governance and operational backbone of the platform,
supporting multi-tenant collaboration while maintaining alignment with EU regulations.

\begin{enumerate}

  \item \textbf{Multi-Layer Access Control System}:  Combines authentication, authorisation, organisation-based management, and permissions. Ensures secure access to services, governance, and role-specific control across the sandbox.

  \item \textbf{User Registration and Onboarding}:  Provides self-registration with email verification, role selection, and organisation assignment. Includes approval workflows and status tracking for pending, approved, rejected, or suspended users.

  \item \textbf{User Status Management}:  Manages the lifecycle of user accounts with clear status transitions. Allows administrators to monitor and adjust access as needed to maintain operational integrity.

  \item \textbf{Advanced User Profile Management}:  Maintains user profiles with personal information, research areas, subscription tiers, and quota allocations. Supports role assignment, permission inheritance, and activity audit trails.

  \item \textbf{Role-Based Access Control (RBAC)}:  Implements a seven-level hierarchy from super administrators to external stakeholders. Enforces least-privilege access and fine-grained permission control \cite{Sandhu1996RBAC}.

  \item \textbf{Multi-Tenant Organisation Management}:  Enables creation and administration of isolated organisations with their own policies. Provides resource quotas, compliance settings, subscription management, and organisation-level analytics dashboards.

  \item \textbf{Dynamic Resource Management and Quotas}:  Allocates CPU, GPU, storage, and API resources with monitoring. Supports allocation, adjustments, usage reporting, and quota reset cycles for fairness and efficiency.

  \item \textbf{Audit and Compliance Logging}:  Captures user actions, system events, security incidents, and data access. Supports GDPR and EU AI Act requirements through evidence collection, lineage tracking, and compliance reporting.

  \item \textbf{Administrative Dashboard}:  Provides a console for user management, approvals, monitoring, and compliance checks. Includes metrics, security dashboards, resource monitoring, and performance analytics.

  \item \textbf{Approval Workflows}:  Manages approvals for onboarding, resource requests, quota increases, and project creation. Supports escalation rules and organisation-specific approval chains.

  \item \textbf{Advanced Security Controls}:  Provides security controls including JWT-based authentication, MFA, encryption in transit and at rest, and strong password policies. Enforces least-privilege principles with permissions and access revocation.

  \item \textbf{System Analytics and Reporting}:  Delivers usage insights, cost analysis, performance trends, and collaboration metrics. Generates compliance reports and supports optimisation of system operations.

  \item \textbf{Scalability and Performance Management}:  Supports performance through distributed architecture, horizontal scaling, caching, and database optimisation. Supports growth in user bases and workloads.

  \item \textbf{Customisation and Extensibility}:  Allows role creation, workflow configuration, and organisation-specific policy settings. Provides integration APIs and a plugin approach for extending platform capabilities.

  \item \textbf{User Experience and Accessibility}: Delivers an intuitive and responsive interface with role-specific views. Supports mobile access, updates, and WCAG-aligned accessibility for inclusive usage across user groups.
\end{enumerate}

\subsection*{Analyses of System \& Administration Features}
The following table provides an analysed summary of the system and administration
features implemented in the AI Sandbox. Each feature is described in terms of
its value, outcome, and relevance to different domains, together with its priority
for implementation. This supports a structured view of governance, compliance, and operational
requirements within the platform.

\begin{table}[H]
\centering
\scriptsize
\caption{Analysed Summary of System \& Administration Features}
\begin{tabular}{p{0.8cm} p{3.8cm} p{5.0cm} p{3.0cm} p{2.0cm}}
\toprule
\textbf{\#} & \textbf{Feature} & \textbf{Value / Outcome} & \textbf{Domain} & \textbf{Priority} \\
\midrule
1  & Multi-Layer Access Control System & Provides secure authentication, authorisation, and organisation-based governance with permissions for controlled access. & All users and administrators & High \\
2  & User Registration \& Onboarding & Supports self-registration, email verification, role selection, and organisation assignment with approval workflows. & All & High \\
3  & User Status Management & Manages account lifecycles (pending, approved, rejected, suspended) ensuring controlled access. & All & High \\
4  & Advanced User Profile Management & Maintains user profiles, subscription tiers, quota allocation, and audit trails for accountability. & Researchers, administrators & High \\
5  & Role-Based Access Control (RBAC) & Implements a seven-level role hierarchy from administrators to external stakeholders, enforcing least-privilege principles. & All & High \\
6  & Multi-Tenant Organisation Management & Enables isolated organisations with quotas, compliance settings, subscription tiers, and analytics dashboards. & Universities, companies & High \\
7  & Dynamic Resource Management \& Quotas & Allocates CPU, GPU, storage, and API limits with monitoring, adjustments, and reset cycles. & Research projects, SMEs & High \\
8  & Audit \& Compliance Logging & Logs user actions, system events, and data access. Supports GDPR and EU AI Act compliance reporting. & Administrators, auditors & High \\
9  & Administrative Dashboard & Provides a console for user management, approvals, monitoring, and compliance checks with metrics. & Administrators & High \\
10 & Approval Workflows & Offers approvals for onboarding, resource allocation, and sensitive operations with escalation rules. & Organisations, administrators & High \\
11 & Advanced Security Controls & Ensures security via JWT, MFA, encryption, permissions, and access revocation. & All & High \\
12 & System Analytics \& Reporting & Delivers usage insights, cost analysis, performance trends, and compliance reports. & Admins, managers & Medium \\
13 & Scalability \& Performance Management & Supports distributed architecture, horizontal scaling, caching, and optimised databases for reliability. & All organisations & High \\
14 & Customisation \& Extensibility & Allows role creation, workflow configuration, and plugin-based extensions. & Administrators, partners & Medium \\
15 & User Experience \& Accessibility & Provides intuitive design with WCAG alignment and mobile-ready access for users. & All end users & High \\
\bottomrule
\end{tabular}
\label{tab:sysadminfeatures}
\end{table}

\subsection{AI Services Features}

Based on the consultations, workshops, and internal R\&D discussions, we have compiled
an initial set of features for the AI Sandbox. These features represent a structured
translation of stakeholder requirements into concrete services. They are not considered
final; instead, they form a baseline that will evolve as more companies and public
organisations join the initiative and contribute additional domain-specific needs.

The current feature set covers benchmarking, experimentation, core AI services,
infrastructure, and compliance. Each feature is briefly described below.

\begin{enumerate}
  \item \textbf{Model Benchmarking Service}:  Provides standardised templates and automated pipelines for model evaluation. Enables SMEs and research groups to benchmark models against datasets and generate comparative reports.

  \item \textbf{Requirement Templates Generation Service}:  Supports transformation of unstructured requirements into structured, compliance-ready formats. Improves consistency, reduces manual effort, and helps identify missing information.

  \item \textbf{Cost Estimation Service}:  Produces cost and resource estimates for projects. Offers breakdowns of compute, storage, and effort requirements to support budgeting and proposal preparation.

  \item \textbf{AI Technique Assessment Service}:  Recommends appropriate AI/ML techniques based on project requirements and available data. Provides guidance for selecting suitable algorithms and workflows.

  \item \textbf{Data Preprocessing Service}:  Supports data cleaning, transformation, and feature engineering. Includes quality reporting to support preparation of datasets for experimentation.

  \item \textbf{Controlled Experiment Execution Service}:  Supports reproducible and secure execution of experiments. Integrates GPU allocation, monitoring, and version control for traceability.

  \item \textbf{Results Dashboard and Reporting Service}:  Provides dashboards and comparative analysis of results. Supports exportable reporting for documentation and decision-making.

  \item \textbf{Text Analysis Service}:  Provides natural language processing tools such as sentiment analysis, classification, and entity recognition. Supports anonymisation and secure processing for sensitive data.

  \item \textbf{Computer Vision Service}:  Offers image classification, object detection, and recognition capabilities. Supports privacy-preserving methods and bias analysis for responsible use.

  \item \textbf{Retrieval-Augmented Generation (RAG) Service}:  Enhances document intelligence through retrieval combined with generation. Supports Q\&A, search, and summarisation with citation tracking.

  \item \textbf{Fine-Tuning Service}:  Enables domain-specific adaptation of AI models with monitoring. Supports customisation of pre-trained models.

  \item \textbf{Speech Processing Service}:  Provides speech-to-text, text-to-speech, and voice analytics. Supports transcription and synthesis.

  \item \textbf{Recommendation Service}:  Implements personalised recommendations using collaborative, content-based, and hybrid methods. Supports privacy through anonymisation and access controls.

  \item \textbf{Time Series Analysis Service}:  Provides forecasting, trend analysis, and anomaly detection for time-dependent data. Supports use cases for finance, manufacturing, and energy.

  \item \textbf{Anomaly Detection Service}:  Detects irregular patterns in structured and unstructured data streams. Uses statistical, ML, and deep learning techniques with alerting.

  \item \textbf{LLM Playground Service}:  Offers an environment to experiment with large language models. Supports text, code, and interactive analysis tasks.

  \item \textbf{GPU Resource Allocation Service}:  Provides on-demand access to GPU resources with allocation and workload monitoring features.

  \item \textbf{Secure Data Storage Service}:  Provides encrypted storage aligned with EU data residency requirements. Integrates access control and audit logging for security and traceability.

  \item \textbf{Experiment Tracking Service}:  Manages the lifecycle of experiments through version control and parameter logging. Supports reproducibility and long-term R\&D tracking.

  \item \textbf{AI Model Vulnerability Scanning Service}:  Scans models and datasets for vulnerabilities, bias, and compliance risks \cite{ENISA2021SMLA,NIST_AIRMf}. Produces reports and remediation guidance.

  \item \textbf{Legal Assistant Service}:  Provides conversational support related to GDPR and EU AI Act compliance. Supports risk assessment and documentation tasks.

  \item \textbf{Data Residency Control Service}:  Allows organisations to manage data storage locations to comply with jurisdictional requirements. Provides monitoring and reporting.

  \item \textbf{Data Anonymization Service}:  Applies anonymisation techniques such as k-anonymity, l-diversity, and t-closeness. Supports GDPR-aligned use of sensitive data while preserving utility \cite{Sweeney2002KAnonymity,Machanavajjhala2007LDiversity,Li2007TCloseness}.

  \item \textbf{AI-Native Anomaly Detection Service}:  Extends anomaly detection with AI/ML and optional LLM integration. Supports monitoring across multiple data types with alerts.

  \item \textbf{Compliance Auditing Service}:  Supports GDPR and EU AI Act compliance checks. Integrates evidence collection, reporting, and risk assessment.

  \item \textbf{AI Security Scanning Service}:  Provides scanning of code, dependencies, and infrastructure. Offers remediation advice to strengthen system security.
\end{enumerate}

\subsection*{Analyses of AI Services Features}
The following table summarises the initial set of AI features identified for the AI
Sandbox through stakeholder consultations, prototype reviews, and internal
brainstorming. Each use case highlights a concrete organisational need, its value,
and priority across domains. These represent a baseline portfolio that will evolve
as more organisations join and contribute additional requirements.

\begin{table}[H]
\centering
\tiny
\caption{Analysed Summary of Use Cases for the AI Sandbox}

\begin{tabular}{p{0.8cm} p{3.5cm} p{5.3cm} p{3.0cm} p{2.0cm}}
\toprule
\textbf{\#} & \textbf{Use Case / Service} & \textbf{Value / Outcome} & \textbf{Domain} & \textbf{Priority}\\
\midrule
1 & Model Benchmarking Service & Compare SME-trained models against industry standards to measure competitiveness & Cross-domain & High \\
2 & Requirement Templates Generation Service & Transform unstructured requirements into structured, compliance-ready documentation & Software engineering, IT services & High \\
3 & Cost Estimation Service & Provide transparent pricing and effort estimation for project proposals using AI analysis & IT services & Medium \\
4 & AI Technique Assessment Service & Recommend suitable AI/ML methods and approaches for specific business or research needs & Cross-domain & High \\
5 & Data Preprocessing Service & Enable automated, standardised, and high-quality preparation of datasets for experiments & All & High \\
6 & Controlled Experiment Execution Service & Ensure reproducible, secure, and monitored execution of experiments with GPU allocation & Research and industry R\&D & High \\
7 & Results Dashboard and Reporting Service & Provide dashboards for monitoring, comparison, and exporting results & All & High \\
8 & Text Analysis Service & Perform NLP including sentiment analysis, classification, and entity recognition & Knowledge-intensive services & Medium \\
9 & Computer Vision Service & Support image classification, detection, and recognition with privacy protection & Manufacturing, healthcare, security & Medium \\
10 & Retrieval-Augmented Generation Service & Deliver Q\&A and document intelligence with citations & Knowledge-intensive services & Medium \\
11 & Fine-Tuning Service & Customise AI models with hyperparameter tuning workflows & AI research, SMEs & Medium \\
12 & Speech Processing Service & Provide speech-to-text and text-to-speech with multilingual support & Public services, education, media & Medium \\
13 & Recommendation Service & Enable personalised recommendations for digital platforms with privacy controls & E-commerce, media & Medium \\
14 & Time Series Analysis Service & Forecast trends and detect anomalies in time-dependent data & Finance, manufacturing, energy & High \\
15 & Anomaly Detection Service & Provide anomaly detection for structured and unstructured data & Cross-domain & High \\
16 & LLM Playground Service & Offer an environment for testing large language models across tasks & Research, education, SMEs & High \\
17 & GPU Resource Allocation Service & Provide GPU resources for scalable AI training and inference & Research, SMEs, IT services & High \\
18 & Secure Data Storage Service & Ensure encrypted, EU-aligned data storage with access controls & All & High \\
19 & Experiment Tracking Service & Support reproducibility with experiment tracking and comparison & Research and innovation projects & High \\
20 & AI Model Vulnerability Scanning Service & Identify vulnerabilities and biases in models and training data & All & High \\
21 & Legal Assistant Service & Provide guidance on GDPR and EU AI Act compliance and related documentation & All & High \\
22 & Data Residency Control Service & Manage data storage locations to meet jurisdictional requirements & All & Medium \\
23 & Data Anonymization Service & Support GDPR-aligned use of sensitive datasets through anonymisation & Healthcare, public services & High \\
24 & AI-Native Anomaly Detection Service & Monitor heterogeneous data streams with AI/LLM integration options & Critical infrastructure, finance & High \\
25 & Compliance Auditing Service & Support GDPR and EU AI Act checks with evidence management & All & High \\
26 & AI Security Scanning Service & Conduct vulnerability scanning across code, infrastructure, and runtime & All & High \\
\bottomrule
\end{tabular}
\label{tab:usecases}
\end{table}

\subsection{Demonstration of Model Usage}

In addition to analysing the openness and licensing aspects, we also
demonstrated how these models could be integrated within the AI Sandbox
environment for different purposes, including text generation, code assistance,
and image-related tasks. The demonstrations showcased workflows
for model selection, execution, and monitoring in a secure and compliant
sandbox setting.

It is important to note that, due to the scope and requirements of this project,
these models were not implemented as production-ready services.
Instead, the demonstrations were designed to illustrate feasibility,
integration workflows, and the potential value for future expansion of the
sandbox platform.

\begin{table}[H]
\centering
\tiny
\caption{Summary of Open and Semi-Open AI Models}
\begin{tabular}{p{3cm} p{3cm} p{2cm} p{3cm} p{4cm}}
\toprule
\textbf{Model / Project} & \textbf{Origin / Maintainer} & \textbf{License Type} & \textbf{Openness Level} & \textbf{Notes} \\
\midrule
Falcon 40B & TII (UAE) & Apache 2.0 & Fully Open & First large-scale model released with full openness. \\
Falcon 7B & TII (UAE) & Apache 2.0 & Fully Open & Efficient lightweight variant for smaller tasks. \\
BLOOM / BLOOM 2 & BigScience / Hugging Face & RAIL & Open weights + data (restricted use) & Multilingual with transparent training data. \\
GPT-J 3.5 & EleutherAI & Apache 2.0 & Fully Open & GPT-3 class model with strong adoption. \\
GPT-NeoX-20B & EleutherAI & Apache 2.0 & Fully Open & Large-scale transformer, widely used. \\
YaLM-100B & Yandex & Apache 2.0 & Fully Open & 100B bilingual model (Russian/English). \\
UL2 / Flan-UL2 & Google & Apache 2.0 & Fully Open & Instruction-tuned, T5-style architecture. \\
Cerebras-GPT & Cerebras & Apache 2.0 & Fully Open & Transparent family of GPT models. \\
StableLM & Stability AI & Apache 2.0 & Fully Open & Open-source focus on chat/instruction. \\
MPT (7B, 30B) & MosaicML & Apache 2.0 & Fully Open & Modular, efficient, long-context models. \\
RedPajama / INCITE & Together AI + partners & Apache 2.0 & Fully Open & Open reproduction of LLaMA dataset. \\
OpenLLaMA & OpenLM Research & Apache 2.0 & Fully Open & LLaMA reproduction with open data. \\
XGen-7B & Salesforce & Apache 2.0 & Fully Open & Efficient transformer for long context. \\
Jais-13B & Inception Institute UAE & Apache 2.0 & Fully Open & Arabic-centric with open weights. \\
OpenHermes & Teknium / community & MIT & Fully Open & Community-driven, chat-focused model. \\
Dolly 3.0 & Databricks & MIT & Fully Open & Instruction-tuned for enterprise tasks. \\
PolyLM (1.7B, 13B) & Shanghai AI Lab & Apache 2.0 & Fully Open & Multilingual, with Asian-language focus. \\
LLaMA 3 & Meta AI & Custom (R/C) & Weights only & Powerful but restricted licensing. \\
Vicuna & LMSYS (LLaMA fine-tune) & Non-commercial & Weights only & Popular for chat, non-commercial license. \\
Mistral 7B v2 & Mistral AI & Apache 2.0 & Weights only (closed data) & Small, efficient, high quality. \\
Gemma 2.0 Flash & Google DeepMind & Custom permissive & Weights only & Lightweight, efficient model. \\
DeepSeek-R1 & DeepSeek AI & Custom & Weights only & Bilingual, Chinese/English focused. \\
Baichuan 2 (7B, 13B) & Baichuan AI & Custom & Weights only & Strong Chinese-language model. \\
ChatGLM & Tsinghua / ZhipuAI & Apache 2.0 & Weights only & Widely used bilingual Chinese model. \\
Grok AI & xAI (Elon Musk) & Unknown & Unclear & Marketed as open, not yet released. \\
\bottomrule
\end{tabular}
\label{tab:openmodels}
\end{table}

\subsection{Hardware Management Features}

The hardware management feature module in the AI Sandbox provides structured access to
computational resources that are essential for AI development, including large language
model training, fine-tuning, and inference. It ensures that researchers and companies
can request, schedule, and monitor hardware usage with transparency in pricing,
compliance, and approval processes. The following features define how resources are
provisioned, optimised, and governed within the sandbox.

\begin{enumerate}

  \item \textbf{HPC Resource Management:} Provides access to high-performance GPU and CPU
  clusters (e.g., NVIDIA A100/H100 GPUs and AMD EPYC CPUs) with detailed specifications,
  availability, and transparent per-hour costs. Supports model training, large-scale
  inference, and data-intensive AI workflows.
  \item \textbf{Storage Resource Management:} Offers high-capacity and high-speed storage
  systems for AI datasets and experiment outputs. Enables secure, redundant, and scalable
  data management with parallel file systems and object storage, ensuring smooth handling
  of large training corpora.
  \item \textbf{Cloud Resource Integration:} Connects to Kubernetes clusters
  across AWS, Azure, and GCP, supporting containerised AI workloads. Supports
  deployment for training, fine-tuning, and distributed inference while
  maintaining EU compliance \cite{KubernetesOverview}.
  \item \textbf{Request Creation Workflow:} Implements a guided multi-step wizard for
  selecting hardware, scheduling usage, and providing justification. Includes cost
  calculation and ensures that requests are tied to approved AI research projects.
  \item \textbf{Approval Workflows:} Supports multi-level administrative approvals for
  resource requests. Validates availability, budget allocations, and organisational
  policies before granting access to GPUs or compute clusters.
  \item \textbf{User Resource Dashboard:} Provides role-based dashboards with
  status of hardware availability, utilisation metrics, and cost estimates. Supports
  monitoring of GPU/CPU usage.
  \item \textbf{Monitoring and Analytics:} Tracks utilisation, experiment throughput,
  and per-project costs. Supports usage optimisation, bottleneck identification, and
  capacity planning for AI experiments.
  \item \textbf{Security and Compliance Controls:} Embeds GDPR compliance, EU AI Act
  alignment, role-based access, encrypted storage, and audit trails. Ensures
  secure and legally compliant use of AI-related resources.
  \item \textbf{Advanced Scheduling:} Enables calendar-based bookings, recurring GPU
  reservations, and priority queues for urgent experiments. Supports conflict handling
  to support fair access and utilisation.
  \item \textbf{Optimisation Features:} Provides recommendations for load
  balancing, energy efficiency, and cost minimisation. Supports organisations in
  managing compute usage across AI research and development.
  \item \textbf{Integration APIs:} Offers REST APIs, webhooks, and SDKs for integrating
  hardware management with external AI pipelines, monitoring systems, and billing
  tools. Supports automation and integration into existing research workflows.
\end{enumerate}

\subsection*{Analysis}

The Hardware Management Module ensures that the AI Sandbox can deliver secure,
scalable, and transparent access to computational resources needed for AI and
large language model development. It integrates GPU/CPU allocation, storage,
cloud integration, scheduling, and monitoring into one governed environment,
enabling organisations to manage costs, optimise performance, and remain
compliant with EU requirements. Table~\ref{tab:hardware} summarises the analysed
features of this module.

\begin{table}[H]
\centering
\tiny
\caption [Hardware Management]{Analysed Hardware Management Features for the AI Sandbox}
\begin{tabular}{p{0.8cm} p{3.5cm} p{5.5cm} p{2.5cm} p{1.5cm}}
\toprule
\textbf{\#} & \textbf{Feature} & \textbf{Value / Outcome} & \textbf{Domain} & \textbf{Priority} \\
\midrule
1 & HPC Resource Management & Provides high-performance GPU/CPU clusters for training, fine-tuning, and inference with transparent costs and availability. & Research, AI development, SMEs & High \\
2 & Storage Resource Management & Ensures scalable, redundant storage for large datasets and experiment outputs with high-speed access. & Data-intensive AI projects & High \\
3 & Cloud Resource Integration & Enables Kubernetes deployments (AWS, Azure, GCP) for distributed AI workloads with EU compliance. & Cross-domain AI research & Medium \\
4 & Request Creation Workflow & Guides users through hardware request and cost estimation tied to project needs. & All & High \\
5 & Approval Workflows & Validates resource requests via multi-level admin approval to enforce governance and fair access. & Research and organisational IT & High \\
6 & User Resource Dashboard & Provides visibility into resource usage, availability, and costs. & All & High \\
7 & Monitoring \& Analytics & Tracks utilisation, throughput, and costs for optimisation, planning, and performance improvements. & Research and R\&D infrastructure & High \\
8 & Security \& Compliance Controls & Embeds GDPR and EU AI Act compliance with encryption, access logs, and audit trails \cite{NIST80053r5}. & All (legal and regulatory) & High \\
9 & Advanced Scheduling & Offers calendar-based bookings, recurring reservations, and priority queuing to optimise utilisation. & Cross-domain & Medium \\
10 & Optimisation Features & Provides recommendations for cost savings, energy efficiency, and balanced utilisation. & All & Medium \\
11 & Integration APIs & Provides APIs and SDKs for integrating hardware management with AI pipelines and billing systems. & Research, IT services & Medium \\
\bottomrule
\end{tabular}
\label{tab:hardware}
\end{table}

\chapter{Architecture and Technology Stack}

\fcolorbox{brand}{panel}{%
  \begin{minipage}{0.97\textwidth}
  \vspace{0.6em}
  \textbf{Highlights}
  \begin{itemize}
    \item Layered and modular architecture of the Sandbox
    \item Secure design with built-in privacy and compliance features
    \item Data flow, integrations, and control mechanisms
    \item Container-based deployment and scalability path
    \item Monitoring, observability, and performance readiness
  \end{itemize}
  \vspace{0.4em}
  \end{minipage}
}

\section*{Overview}
This part of the deliverable defines the sandbox’s high-level architecture and the technology stack that enables secure, compliant AI experimentation. It specifies the system’s key components and interfaces, the deployment model, and the mechanisms that ensure privacy, security, and regulatory alignment (GDPR and EU AI Act). The content is kept practical and implementation-oriented to reflect the current prototype.

\section{Architecture and Key Components}
\label{sec:arch-and-components}

\subsection{Overview}
The AI Sandbox prototype was implemented using a layered and microservices-based (MSA) architecture \cite{FowlerMicroservicesDefinition,FowlerMicroservicePremium,FowlerMicroservicePrereq}. Each layer has a defined responsibility and interacts with neighbouring layers through stable APIs. This combination supports modularity, scalability, and maintainability, and it is realised in the working demo.

\begin{figure}[htbp]
  \centering
  \includegraphics[width=\textwidth]{\detokenize{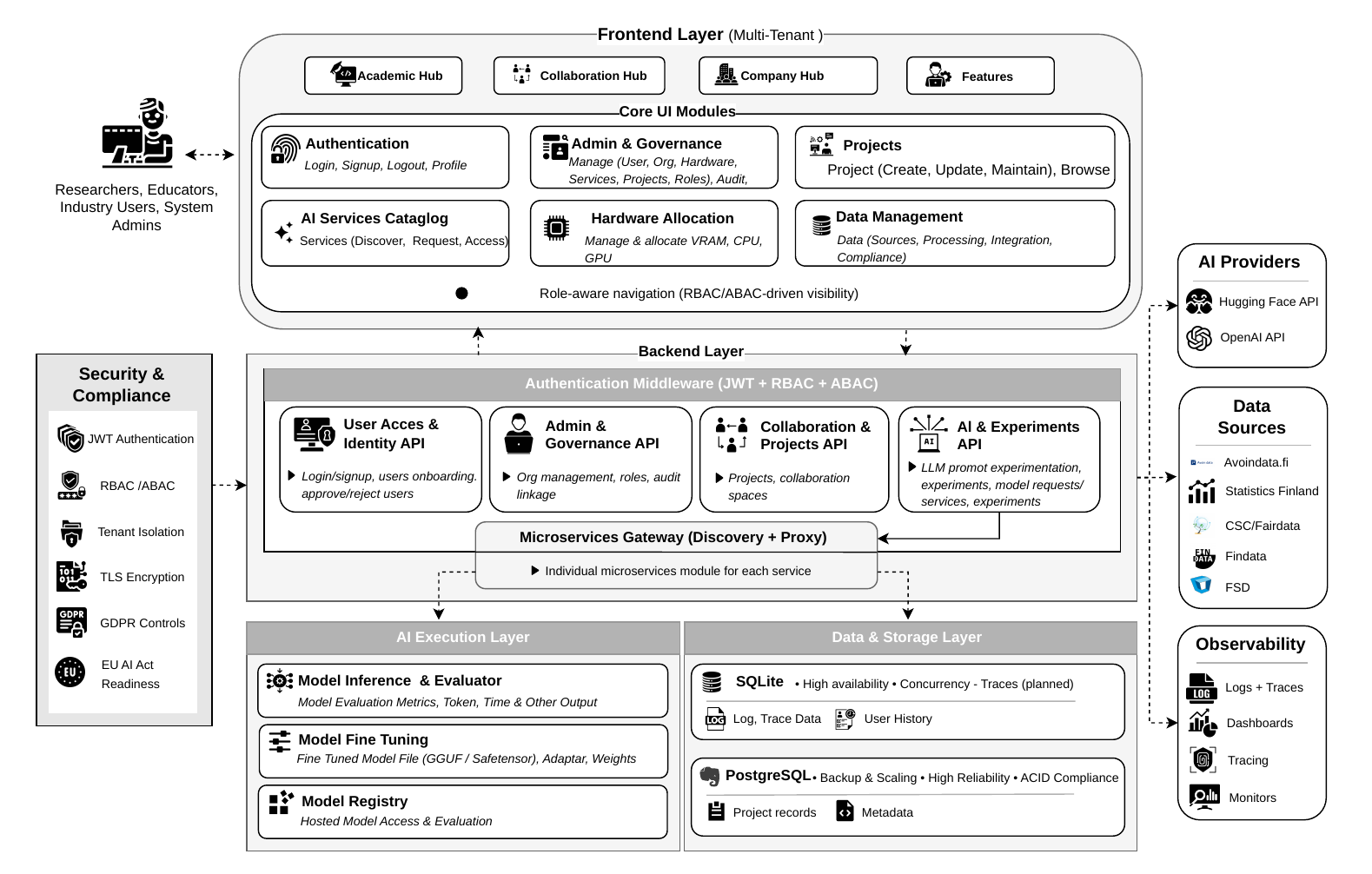}}
  \caption[AI Sandbox high-level architecture]{AI Sandbox high-level architecture with layered design and microservices-based components.}
  \label{fig:sandbox-arch}
\end{figure}

Figure~\ref{fig:sandbox-arch} provides the structural view. User roles interact with the \textbf{frontend application} (Next.js, React/TypeScript, Tailwind CSS, JWT-based authentication) \cite{NextJSAppRouter,ReactDocs}. The frontend communicates with \textbf{backend microservices} (Express.js on Node.js) that expose REST APIs \cite{Express4API,RFC9110,RFC9111}. These services orchestrate \textbf{AI capabilities} such as anomaly detection, preprocessing pipelines, security scanning, compliance auditing, and an LLM playground. All interactions are governed by a \textbf{data and security layer} that enforces encryption, access control (RBAC and ABAC), tenant isolation, and audit logging.

\subsection{Architectural Style and Principles}
The prototype follows a layered model and an MSA style in parallel:
\begin{itemize}
  \item \textbf{Layered modularity}: Clear separation of frontend, backend services, AI services, and data/security.
  \item \textbf{Microservices orientation}: Independent services communicate over REST (\texttt{/api/auth}, \texttt{/api/projects}, \texttt{/api/ai-services}, \texttt{/api/hardware}, etc.).
  \item \textbf{Cloud-native}: Containerised services with a migration path from Docker Compose (prototype) to Kubernetes (production).
  \item \textbf{Security by design}: TLS 1.2+ (TLS 1.3 preferred), AES-256 at rest, RBAC/ABAC \cite{Sandhu1996RBAC,NIST800162}, MFA where required, tenant isolation, and continuous audit trails \cite{RFC8446,RFC5246,FIPS197}.
  \item \textbf{Compliance integration}: GDPR and EU AI Act guardrails embedded in workflows; alignment with Finnish regulations (Findata, THL, FSD) when applicable.
  \item \textbf{Role-centric UX}: Dashboards tailored for Super Admin, Research Admin, Researchers, University Faculty, and Company Admins.
  \item \textbf{Scalability and extensibility}: Horizontal scaling, database migration from SQLite to PostgreSQL, and pluggable AI service integrations.
\end{itemize}

\subsection{Layers and Interfaces}
The following sublayers and interfaces make the architecture operational. Each description links the user-facing needs to the underlying services and controls.

\subsubsection{User Interface}
Role-specific portals reflect permissions and tasks.
\begin{itemize}
  \item \textbf{Landing and Authentication}: Entry point, login, registration, session handling.
  \item \textbf{Dashboards}: Personalised views for students, researchers, and companies with notifications and quick access to services.
  \item \textbf{AI Services Catalogue}: Browse and request access to registered services (more than 24 available in scope during prototyping, with a broader catalogue defined in the requirements chapter).
  \item \textbf{Admin Console}: Configure services, monitor usage, manage users, and enforce policies.
\end{itemize}

\subsubsection{Backend and Microservices}
Stateless services expose stable REST APIs and enforce access policies.
\begin{itemize}
  \item \textbf{Authentication and Authorisation}: JWT and JOSE for tokens, RBAC/ABAC checks, optional MFA \cite{RFC7519,RFC7515,RFC7516,RFC7518}.
  \item \textbf{Core APIs}: \texttt{/api/auth}, \texttt{/api/users}, \texttt{/api/projects}, \texttt{/api/ai-services} (catalogue, execute, results), \texttt{/api/hardware} (requests and approvals).
  \item \textbf{Operational Services}: Access request workflow, admin management, file upload, notifications.
\end{itemize}

\subsubsection{Data and Storage}
Persistent stores and data utilities support projects and auditability.
\begin{itemize}
  \item \textbf{Prototype store}: SQLite for low-ops persistence with fast bindings.
  \item \textbf{Production target}: PostgreSQL for concurrency, migrations, backup, and high availability.
  \item \textbf{Core entities}: Users and roles, projects, AI services, hardware requests, audit logs, notifications, file metadata.
\end{itemize}

\subsubsection{Security and Compliance}
Controls are applied at every boundary and recorded for audit.
\begin{itemize}
  \item \textbf{Access control}: RBAC and ABAC with least privilege and separation of duties.
  \item \textbf{Protection measures}: TLS 1.2+ in transit (TLS 1.3 preferred), AES-256 at rest, input validation, rate limiting, security headers.
  \item \textbf{Compliance}: GDPR mechanisms (consent, subject rights, DPIA, breach notification), EU AI Act readiness (risk classification, human oversight, transparency), and national guidelines where relevant.
  \item \textbf{Monitoring and evidence}: Immutable audit logs, health checks, and operational metrics.
\end{itemize}

\subsubsection{External Integrations}
The platform connects to AI providers and public data sources through controlled interfaces.
\begin{itemize}
  \item \textbf{AI providers}: Hugging Face API, OpenAI API, and custom services for anomaly detection, preprocessing, security scanning, and compliance auditing.
  \item \textbf{Data sources}: Avoindata.fi, Statistics Finland, THL Sotkanet, Findata, FSD, EU Open Data Portal, ELIXIR, and CSC resources as applicable.
  \item \textbf{Support services}: Email notifications, file storage backends, analytics, and monitoring endpoints.
\end{itemize}

\subsection{Implemented Prototype Footprint}
The demo validates the architecture in a realistic setup and defines a clear route to production.
\begin{itemize}
  \item \textbf{Frontend}: Next.js 14.2.5 with React/TypeScript and Tailwind CSS, running on port 3000.
  \item \textbf{Backend}: Express.js on Node.js 18+, running on port 3001, exposing the control plane and AI service endpoints.
  \item \textbf{Persistence}: SQLite in the prototype with a planned migration to PostgreSQL in production.
  \item \textbf{AI services}: Anomaly detection, data preprocessing pipelines, security scanner, compliance auditor, and LLM playground, orchestrated through the backend APIs.
  \item \textbf{Security}: JWT-based authentication, RBAC and ABAC, tenant isolation, audit logging, and input validation.
  \item \textbf{Deployment path}: Docker and Docker Compose in development and staging, with Kubernetes and PostgreSQL as production targets \cite{KubernetesOverview,KubernetesMultiTenancy,DockerMultistage}.
\end{itemize}

\section{Security and Compliance}
\label{sec:security-compliance}

\begin{figure}[htbp]
  \centering
  \includegraphics[width=0.86\linewidth]{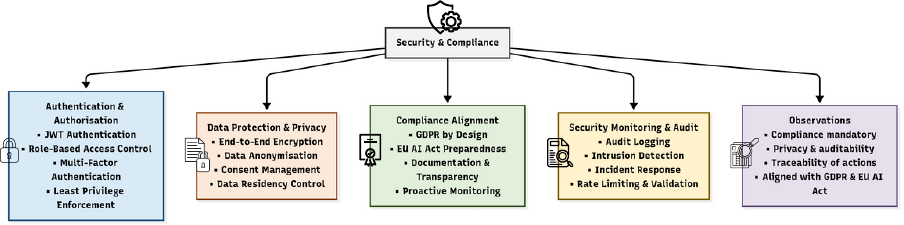}
  \caption[Security and Compliance]{Security and Compliance: control families applied across layers and interfaces (AI-assisted illustration).}
  \label{fig:SecurityandCompliance}
\end{figure}

Security and compliance operate as a cross-cutting layer over the architecture in Figure~\ref{fig:sandbox-arch}. Figure~\ref{fig:SecurityandCompliance} summarises the control set implemented in the prototype: (i) \emph{Authentication \& Authorisation}, (ii) \emph{Data Protection \& Privacy}, (iii) \emph{Compliance Alignment}, (iv) \emph{Security Monitoring \& Audit}, and (v) \emph{Observations/Outcomes}. These controls are exercised along the request path described in Figure~\ref{fig:DataFlowandIntegration}—from the UI, through service APIs, to data stores and external providers—and are evidenced by audit logs, consent records, and configuration policies.

\subsection*{Authentication \& Authorisation (left block in Figure~\ref{fig:SecurityandCompliance})}
\begin{itemize}
  \item \textbf{JWT/JOSE}: Signed (JWS) and, where required, encrypted (JWE) tokens; short-lived access tokens with refresh \cite{RFC7519,RFC7515,RFC7516,RFC7518,Sandhu1996RBAC,NIST800162}.
  \item \textbf{RBAC}: Role-based permissions aligned with system roles (student, researcher, company, organisation admin, system admin).
  \item \textbf{ABAC}: Attribute rules (e.g., organisation, project, dataset class) for fine-grained policies.
  \item \textbf{MFA}: Prompts for privileged actions (admin configuration, dataset upload/approval).
  \item \textbf{Least Privilege \& Separation of Duties}: Requesters cannot approve their own access; services run with minimal scopes.
\end{itemize}

\subsection*{Data Protection \& Privacy (second block)}
\begin{itemize}
  \item \textbf{Encryption in Transit}: TLS~1.3 (fallback to 1.2 only if necessary) for client↔service and service↔service traffic.
  \item \textbf{Encryption at Rest}: AES-256 for databases and backups; keys retained within EU\/EEA jurisdictions \cite{RFC8446,RFC5246,FIPS197}.
  \item \textbf{Anonymisation/Pseudonymisation}: GDPR-aligned techniques for PII before non-production use or cross-tenant sharing.
  \item \textbf{Consent Management}: Explicit consent for uploads/sharing with verifiable, time-stamped records.
  \item \textbf{Data Residency Control}: Policies and storage classes that enforce regional location requirements.
\end{itemize}

\subsection*{Compliance Alignment (centre block)}
\begin{itemize}
  \item \textbf{GDPR by Design}: Subject access/erasure workflows, purpose limitation, data minimisation, comprehensive auditability.
  \item \textbf{EU AI Act Preparedness}: Service risk classification (minimal/limited/high), human-oversight checkpoints, robustness testing, transparency artefacts.
  \item \textbf{Documentation \& Transparency}: Model/dataset metadata, decision logs, configuration snapshots, usage histories retained for audits.
  \item \textbf{National Guidance (Finland)}: Alignment with Findata/THL/FSD for health and social-science datasets where applicable.
\end{itemize}

\subsection*{Security Monitoring \& Audit (fourth block)}
\begin{itemize}
  \item \textbf{Audit Logging}: Immutable, time-stamped records of user and system actions; cryptographically verifiable where required.
  \item \textbf{Intrusion/Anomaly Detection}: Alerts for unusual authentication patterns, API misuse, and denial-of-service attempts.
  \item \textbf{Incident Response}: Playbooks for detection, escalation, containment, recovery, and post-incident review.
  \item \textbf{Protective Controls}: Rate limiting, schema-based input validation, and secure headers to mitigate common web threats.
\end{itemize}

\section{Data Flow and Integration}
\label{sec:data-flow-integration}

Building on the control families in Section~\ref{sec:security-compliance} and the structural view in Figure~\ref{fig:sandbox-arch}, this section shows \emph{where} those controls are exercised along the request path. Figure~\ref{fig:DataFlowandIntegration} summarises the data-flow and integration view and should be read together with the security overlay in Figure~\ref{fig:SecurityandCompliance}.

\subsection*{Primary Request–Response Cycle}
The core cycle has eight stages; each stage cites the principal control(s) applied:
\begin{enumerate}
  \item \textbf{User Request}: A user initiates an action via the frontend (e.g., login, project update, AI service execution). \emph{Entry checks}: input validation and basic rate limiting.
  \item \textbf{Authentication}: The request passes through JWT/JOSE middleware for token validation and session management; MFA is prompted for privileged actions. \emph{Controls}: TLS~1.3, short-lived tokens with refresh.
  \item \textbf{Authorisation}: RBAC/ABAC verifies that the caller has the privileges required for the requested action. \emph{Controls}: least privilege and separation of duties.
  \item \textbf{Business Logic}: The backend service applies workflow rules (e.g., access request approval, service execution parameters) and records decisions. \emph{Controls}: structured audit events.
  \item \textbf{Data Access}: Validated queries execute against the database. \emph{Controls}: encryption at rest (AES-256), residency policies, consent/anonymisation where applicable, immutable audit logs.
  \item \textbf{External Integration}: When required, the service calls third-party APIs (e.g., Hugging Face, OpenAI) in a controlled and logged manner. \emph{Controls}: outbound allow-lists, scoped tokens/keys, provider SLAs/DPA alignment, latency/error monitoring.
  \item \textbf{Response}: The service returns a structured result (JSON/HTML) to the frontend. \emph{Controls}: data minimisation, appropriate cache headers.
  \item \textbf{Notification}: The system emits user notifications (email/dashboard) and operational/audit events. \emph{Controls}: signed events and retention policies.
\end{enumerate}

\subsection*{API Integration Points}
The platform uses well-defined interfaces for internal and external communication:
\begin{itemize}
  \item \textbf{Internal REST APIs}: Dashboards communicate with microservices via stable endpoints (\texttt{/api/auth}, \texttt{/api/users}, \texttt{/api/projects}, \texttt{/api/ai-services}, \texttt{/api/hardware}).
  \item \textbf{Database Access Layer}: Secure, abstracted access to SQLite in the prototype and PostgreSQL in production; migrations and audit hooks included.
  \item \textbf{External AI APIs}: Hugging Face and OpenAI for model access and execution, called through gateway functions with scoped credentials.
  \item \textbf{Notification APIs}: Email and alerting services to surface important system events to users and admins \cite{HFInference,HFProviders,OpenAIAPI}.
\end{itemize}

\subsection*{Integration with External Services}
Third-party services are integrated with explicit guardrails:
\begin{itemize}
  \item \textbf{Hugging Face API}: Access and deploy pre-trained models within scoped rate limits and logs.
  \item \textbf{OpenAI API}: GPT-based LLM capabilities via audited requests; optional EU-hosted endpoints when required.
  \item \textbf{Monitoring Services}: Metrics and error rates for observability across external calls.
  \item \textbf{File Storage Services}: Secure handling of user datasets with presigned access and retention controls.
\end{itemize}

\subsection*{Observations}
\begin{itemize}
  \item \textbf{Traceability}: Requests are traceable end-to-end via structured audit events and logs.
  \item \textbf{Defence in Depth}: Authentication, authorisation, validation, and encryption appear at multiple stages.
  \item \textbf{Integration Readiness}: Standard APIs enable interaction with external providers without weakening compliance.
  \item \textbf{Scalable Flows}: The modular design supports parallel requests and external calls, with clear points for caching and back-pressure.
\end{itemize}

\begin{figure}[htbp]
  \centering
  \includegraphics[width=0.75\linewidth]{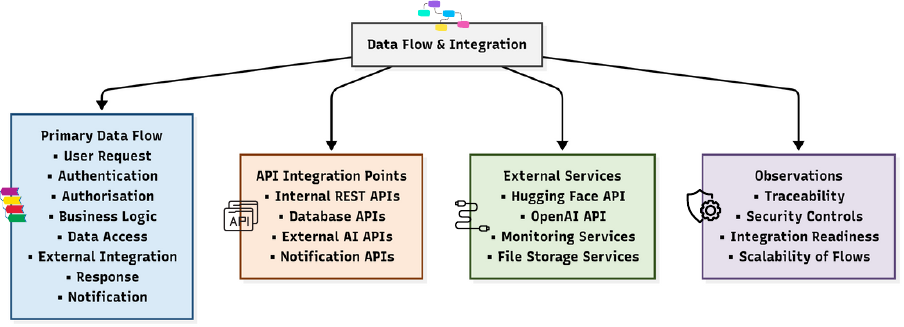}
  \caption[Data Flow \& Integration]{Data Flow and Integration across stages and interfaces (to be read with Figures~\ref{fig:sandbox-arch} and~\ref{fig:SecurityandCompliance}). (AI-assisted illustration)}
  \label{fig:DataFlowandIntegration}
\end{figure}

\section{Deployment Architecture}
\label{sec:deployment-architecture}

\begin{figure}[htbp]
  \centering
  \includegraphics[width=0.9\linewidth]{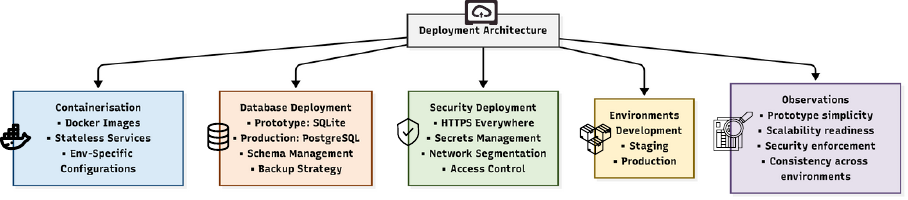}
  \caption[Deployment Architecture]{Deployment Architecture across development, staging, and production (to be read with Figures~\ref{fig:sandbox-arch} and~\ref{fig:DataFlowandIntegration}).}
  \label{fig:DeploymentArchitecture}
\end{figure}

Following the controls in Section~\ref{sec:security-compliance} and the request path in Figure~\ref{fig:DataFlowandIntegration}, this section describes \emph{where} the system runs and \emph{how} it is packaged, configured, and promoted across environments. Figure~\ref{fig:DeploymentArchitecture} summarises the baseline deployment model that was used for the prototype and the planned path to production.

\subsection*{Containerisation}
All frontend, backend, and supporting services are packaged as containers to keep build and runtime consistent:
\begin{itemize}
  \item \textbf{Docker images}: Multi-stage builds produce small, reproducible images for the Next.js frontend and Express-based services.
  \item \textbf{Stateless services}: Containers are designed to be stateless; persistent state is kept in the database or object storage, simplifying scaling and failover.
  \item \textbf{Configuration profiles}: Separate compose files and environment files for development, staging, and production ensure predictable rollouts.
\end{itemize}

\subsection*{Database Deployment}
The data layer follows a clear prototype-to-production migration path:
\begin{itemize}
  \item \textbf{Prototype}: SQLite3 for lightweight local persistence without a dedicated server.
  \item \textbf{Production target}: PostgreSQL (or equivalent) for concurrency, migrations, backup/restore, and high availability.
  \item \textbf{Schema management}: Version-controlled migrations are applied as part of the release pipeline to keep environments aligned.
  \item \textbf{Backup strategy}: Automated backups with recurrent recovery tests to verify restore procedures.
\end{itemize}

\subsection*{Security Deployment}
Security controls from Section~\ref{sec:security-compliance} are enforced at runtime:
\begin{itemize}
  \item \textbf{TLS everywhere}: TLS~1.3 for all client–service and service–service traffic; certificates managed per environment.
  \item \textbf{Secrets management}: In the prototype, environment variables are used with rotation; production adopts a secrets manager (cloud KMS/Vault) and scoped service identities.
  \item \textbf{Network segmentation}: Frontend, services, and databases are placed on separate networks/VPC segments with strict inbound/outbound rules.
  \item \textbf{Access control}: Admin panels and APIs require authenticated, authorised access; audit events are shipped to central logs as part of the deployment.
\end{itemize}

\subsection*{Environments}
The system is promoted through three stages, keeping configuration and evidence consistent:
\begin{itemize}
  \item \textbf{Development}: Local containerised setup with debug tools, seeded data, and permissive logging to speed iteration.
  \item \textbf{Staging}: Mirrors production topology with anonymised datasets for pre-release testing and performance checks.
  \item \textbf{Production}: Hardened environment with strict policies, high availability targets, and continuous compliance monitoring.
\end{itemize}

\subsection*{Observations}
This deployment approach supports the architectural goals:
\begin{itemize}
  \item \textbf{Prototype simplicity}: Fast iteration with Docker and SQLite, aligned with the flows in Figure~\ref{fig:DataFlowandIntegration}.
  \item \textbf{Scalability readiness}: Defined migration paths to PostgreSQL and Kubernetes for multi-service scaling.
  \item \textbf{Built-in security}: TLS, segmented networks, and secret handling reflect the control set in Section~\ref{sec:security-compliance}.
  \item \textbf{Consistency}: Reproducible builds and environment-specific profiles reduce drift and increase confidence in releases.
\end{itemize}

\section{Performance and Scalability}
\label{sec:performance-scalability}

\begin{figure}[htbp]
  \centering
  \includegraphics[width=0.9\linewidth]{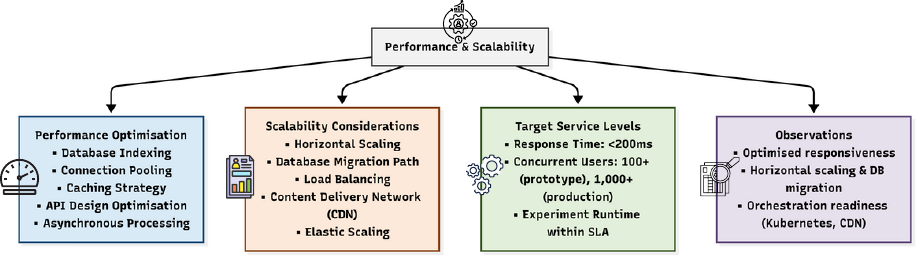}
  \caption[Performance \& Scalability]{Performance levers and scalability path (read with Figures~\ref{fig:DataFlowandIntegration} and~\ref{fig:DeploymentArchitecture}). (AI-assisted illustration)}
  \label{fig:PerformanceandScalability}
\end{figure}

Building on the runtime model in Section~\ref{sec:deployment-architecture} and the request path in Figure~\ref{fig:DataFlowandIntegration}, this section describes how the prototype achieves low latency today and how it scales to production. Figure~\ref{fig:PerformanceandScalability} groups the main levers applied at the API, data, and delivery layers.

\subsection*{Performance Optimisation}
The system applies practical techniques to reduce latency and stabilise throughput:
\begin{itemize}
  \item \textbf{Database indexing}: Targeted indexes speed up common lookups (e.g., users, projects, access requests).
  \item \textbf{Connection pooling}: Reuses DB connections to avoid setup overhead under bursty load (especially after migrating from SQLite to PostgreSQL).
  \item \textbf{Caching strategy}: Caches frequently accessed data (service metadata, session context) to reduce round trips.
  \item \textbf{API design}: REST endpoints use clear HTTP semantics, compact JSON, pagination, and conditional requests (ETag/If-None-Match) where applicable \cite{RFC9110,RFC9111}.
  \item \textbf{Asynchronous processing}: Moves non-critical work (email, log shipping, report generation) off the request path.
  \item \textbf{Asset optimisation}: Tree-shaking and code-splitting in Next.js, gzip/brotli compression, and long-lived cache headers for static assets.
\end{itemize}

\subsection*{Scalability Considerations}
The architecture scales horizontally first, with clear paths for stateful services:
\begin{itemize}
  \item \textbf{Horizontal scaling}: Frontend and backend run as multiple container replicas behind a load balancer.
  \item \textbf{Database migration path}: SQLite (prototype) to \textbf{PostgreSQL} (production) for concurrency, transactions, and HA.
  \item \textbf{Load balancing}: Reverse proxy distributes traffic evenly and supports zero-downtime rollouts.
  \item \textbf{CDN}: Static assets (JS/CSS/images) are served via a content delivery network to cut latency.
  \item \textbf{Elastic scaling}: Kubernetes orchestration allows autoscaling and pod rescheduling as demand changes.
  \item \textbf{Queues \& cache (when needed)}: Redis for sessions/caching and a lightweight queue (e.g., RabbitMQ/NATS) for background jobs.
\end{itemize}

\subsection*{Target Service Levels}
While still a prototype, we define measurable objectives to guide tuning and capacity planning:
\begin{itemize}
  \item \textbf{Response time}: $<$\,200\,ms at the 95th percentile for typical API calls under normal load.
  \item \textbf{Concurrent users}: 100+ concurrent users demonstrated in prototype; production target $\geq$\,1{,}000 concurrent sessions.
  \item \textbf{Workflow latency}: Benchmarking, anonymisation, and similar jobs complete within SLA thresholds (seconds to minutes, depending on payload).
\end{itemize}

\subsection*{Observations}
\begin{itemize}
  \item The prototype delivers responsive behaviour on lightweight infrastructure through indexing, caching, and async work.
  \item Scaling focuses on horizontal replication and migration to enterprise-grade databases, with Kubernetes enabling elastic capacity.
  \item CDN delivery and controlled rollouts (Section~\ref{sec:deployment-architecture}) prepare the system for production and regional expansion.
\end{itemize}

\section{Monitoring and Observability}
\label{sec:monitoring-observability}

\begin{figure}[htbp]
  \centering
  \includegraphics[width=0.8\linewidth]{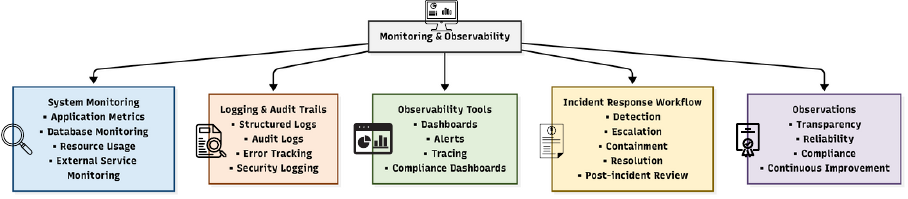}
  \caption[Monitoring \& Observability]{Monitoring and observability across metrics, logs, traces, and compliance views (read with Figures~\ref{fig:DataFlowandIntegration} and~\ref{fig:DeploymentArchitecture}) (AI-assisted illustration). }
  \label{fig:MonitoringandObservability}
\end{figure}

Building on the runtime model in Section~\ref{sec:deployment-architecture} and the request path in Figure~\ref{fig:DataFlowandIntegration}, this section explains how the prototype is instrumented for reliability, trust, and compliance. Instrumentation spans APIs, data stores, infrastructure, and external integrations, with evidence captured for the control set in Section~\ref{sec:security-compliance}. Figure~\ref{fig:MonitoringandObservability} provides the overview.

\subsection*{System Monitoring}
The platform monitors key resources and services to ensure availability and performance:
\begin{itemize}
  \item \textbf{Application metrics}: Response times, error rates, and throughput for backend services.
  \item \textbf{Database monitoring}: Query latency, connection-pool utilisation, and storage growth.
  \item \textbf{Resource usage}: CPU, memory, container health, and node capacity to prevent saturation.
  \item \textbf{External-service monitoring}: Latency and error responses for Hugging Face and OpenAI calls to detect integration failures early.
\end{itemize}

\subsection*{Logging and Audit Trails}
Structured logging and audit trails provide traceability and compliance evidence (see Section~\ref{sec:security-compliance}):
\begin{itemize}
  \item \textbf{Structured logs}: JSON-formatted logs from all services for parsing and correlation.
  \item \textbf{Audit logs}: Immutable records of user actions (logins, uploads, service requests) for forensic analysis.
  \item \textbf{Error tracking}: Centralised capture of exceptions and failure events for debugging.
  \item \textbf{Security logging}: Signals for suspicious logins, denied access, and abuse patterns.
\end{itemize}

\subsection*{Observability Tools}
Operational visibility is delivered through dashboards, alerts, and traces:
\begin{itemize}
  \item \textbf{Dashboards}: Real-time views of SLIs/SLOs for service health and capacity.
  \item \textbf{Alerts}: Policy-driven notifications for SLA breaches, downtime, and anomalies.
  \item \textbf{Tracing}: Distributed traces connect frontend requests to backend and database operations.
  \item \textbf{Compliance dashboards}: Summaries of GDPR and EU AI Act obligations (consent, access/erasure requests, audit completeness).
\end{itemize}

\subsection*{Incident Response Workflow}
The system follows predefined steps to handle incidents end to end:
\begin{enumerate}
  \item Detection through monitoring or alert.
  \item Automatic escalation to system administrators.
  \item Containment of the affected service or environment.
  \item Resolution via runbooks with change records.
  \item Post-incident review to improve controls and prevent recurrence.
\end{enumerate}

\subsection*{Observations}
\begin{itemize}
  \item \textbf{Transparency}: Significant events are logged and auditable across the request path in Figure~\ref{fig:DataFlowandIntegration}.
  \item \textbf{Reliability}: Proactive alerts reduce time to detect and time to recover.
  \item \textbf{Compliance}: Logs and dashboards provide evidence for GDPR and EU AI Act requirements.
  \item \textbf{Continuous improvement}: Incident reviews feed upgrades to deployment and security controls (Sections~\ref{sec:deployment-architecture} and~\ref{sec:security-compliance}).
\end{itemize}

\section{Technology Stack}
\label{sec:technology-stack-production}
Table~\ref{tab:tech-stack-prod-portrait} summarises the stack used in the demo prototype, grouped by frontend, backend, data layer, containers/orchestration, security/compliance, integrations, and operations/CI/CD. Each row lists the tool and version, its role, a production-readiness signal (\ready\= ready; \cond\ = usable with conditions), and practical alternatives. To move beyond the prototype: keep Express or switch to Fastify/NestJS for higher throughput; migrate SQLite to \textbf{PostgreSQL} and add \textbf{Redis} plus a queue as workloads grow; retain JWT/JOSE but add \textbf{OIDC} and a secret manager; standardise observability on \textbf{OpenTelemetry} with Prometheus/Grafana, ELK/OpenSearch, and Sentry; adopt \textbf{Kubernetes} (or start with a managed PaaS) \citep{OTelSpec,OTLP,PrometheusOverview}; and add an EU-hosted inference gateway (vLLM/TGI) for the AI layer. Typical migration steps are SQLite\(\rightarrow\)PostgreSQL, Docker Compose\(\rightarrow\)Kubernetes, JWT\(\rightarrow\)OIDC/SSO, and basic logs\(\rightarrow\)OpenTelemetry/Prometheus/ELK.

\setlength{\tabcolsep}{6pt}
\renewcommand{\arraystretch}{1}
{\scriptsize
\begin{longtable}{P{0.1\linewidth} P{0.2\linewidth} P{0.08\linewidth} P{0.2\linewidth} P{0.06\linewidth} P{0.30\linewidth}}

\caption[Technology Stack]{Technology Stack — production readiness and credible alternatives. Legend: \ready\;ready, \cond\;usable with conditions.}
\label{tab:tech-stack-prod-portrait} \\
\toprule
\textbf{Category} & \textbf{Technology / Service} & \textbf{Version} & \textbf{Purpose (what we implemented in the demo)} & \textbf{Prod} & \textbf{Alternatives when scaling / hardening} \\
\midrule
\endfirsthead

\multicolumn{6}{l}{\footnotesize \textbf{AI Sandbox Technology Stack — continued}}\\
\toprule
\textbf{Category} & \textbf{Technology / Service} & \textbf{Version} & \textbf{Purpose (what we implemented in the demo)} & \textbf{Prod} & \textbf{Alternatives when scaling / hardening} \\
\midrule
\endhead

\midrule
\multicolumn{6}{r}{\footnotesize Continued on next page} \\
\endfoot

\bottomrule
\endlastfoot

\rowcolors{2}{gray!10}{white}

Frontend — Core Framework
  & Next.js + React + TypeScript
  & 14.2.5 / 18.3.1 / 5.9.2
  & SSR/SSG app shell and routing; component model with strong typing across UI and API types.
  & \ready
  & Remix; SvelteKit; Vue~+~Nuxt; Astro (islands). \\

Frontend — Runtime
  & Node.js (dev/build)
  & 18+
  & Local dev server and build pipeline for Next.js.
  & \ready
  & — \\

Styling \& UI
  & Tailwind CSS; Lucide React; Framer Motion; Class Variance Authority; Tailwind Merge; Radix UI (Slider)
  & 3.4.10; 0.451.0; 11.2.12; 0.7.1; 2.6.0; 1.3.6
  & Utility-first styling, iconography, animations, robust variant patterns, and accessible controls.
  & \ready
  & MUI; shadcn/ui; Chakra UI. \\

State / Server Cache
  & React Context; TanStack React Query
  & Built-in; 5.48.0
  & Auth/session prefs and server-state caching with stale-while-revalidate behaviour.
  & \ready
  & Redux Toolkit (complex global state); Zustand (lightweight). \\

Notifications
  & React Toastify
  & 11.0.5
  & User feedback and alert toasts across flows.
  & \ready
  & Notistack; Radix Toast. \\

Data Visualisation
  & Recharts
  & 2.15.4
  & Role/usage metrics and service results charts.
  & \cond
  & ECharts; Vega-Lite; Plotly (heavy interactivity). \\

Frontend Dev Tools
  & PostCSS; Autoprefixer; ESLint
  & 8.4.39; 10.4.19; 8.57.0
  & CSS post-processing, vendor prefixing, linting and code quality.
  & \ready
  & Biome (lint/format); Prettier (format). \\

Backend — Core Framework
  & Express.js on Node.js, FastAPI on Python
  & 4.19.2 / 18+
  & REST APIs for auth, users, projects, AI services; control-plane endpoints.
  & \ready
  & Fastify (higher perf); NestJS (DI/modular); Spring Boot; .NET Minimal APIs; Go (Fiber/Gin). \\

Backend — Language
  & TypeScript, Python
  & 5.4.5
  & Strong typing, DTOs, and runtime validation pairing with Zod.
  & \ready
  & — \\

Auth \& Security
  & JSON Web Tokens (jsonwebtoken); JOSE
  & 9.0.2; 5.2.4
  & Token-based auth, JWS/JWE; stateless sessions.
  & \ready
  & OIDC/SSO (Keycloak, Auth0); PASETO tokens. \\

HTTP Security
  & Helmet.js; express-rate-limit; CORS
  & 7.1.0; 7.3.1; 2.8.5
  & Security headers, DDoS/request throttling, cross-origin policy.
  & \ready
  & WAF / API Gateway policies. \\

Input Validation
  & Zod
  & 4.0.17
  & Schema validation for request bodies/params.
  & \ready
  & class-validator (NestJS); Ajv (JSON Schema). \\

Middleware \& Util
  & Morgan; Cookie Parser; Express Session; Multer; UUID; Axios
  & 1.10.0; 1.4.7; 1.18.2; 2.0.2; 11.1.0; 1.7.2
  & HTTP logging; cookies; sessions; uploads; IDs; outbound HTTP client.
  & \ready
  & pino (structured logs); busboy/formidable (uploads); ky/got (HTTP). \\

Database — Engines
  & SQLite (sqlite3; better-sqlite3; sqlite)
  & 5.1.7; 12.4.1; 5.1.1
  & Prototype relational store; low-ops single-file DB with fast bindings.
  & \cond
  & PostgreSQL (prod target); CockroachDB (distributed SQL); MySQL/MariaDB. \\

Database — CSV / ETL
  & fast-csv; csv-parser
  & 5.0.5; 3.2.0
  & Import/export utilities for datasets and results.
  & \ready
  & Apache Arrow; DuckDB for analytics ETL. \\

ORM / Data Access
  & Direct driver + better-sqlite3
  & —
  & Thin data-access layer for CRUD; simple migrations.
  & \cond
  & Prisma; Drizzle ORM; TypeORM; Knex (query builder). \\

Caching / Queues
  & —
  & —
  & Not required in prototype.
  & \cond
  & Redis (cache/session); RabbitMQ; NATS; Kafka (event streaming). \\

Backend Dev \& Build
  & TSX; ts-node-dev; dotenv; ESLint
  & 4.20.5; 2.0.0; 16.4.5; 8.57.0
  & TS execution, live-reload, env config, linting.
  & \ready
  & SWC/tsx-node; env vaults. \\

Containers \& Base Images
  & Docker; Docker Compose; node:18-alpine
  & 20.10+; 2.0+; 18-alpine
  & Containerised frontend/backend; multi-service dev with Compose.
  & \ready
  & Podman; Buildx/BuildKit; Bazel \texttt{rules\_docker}. \\

Container Config
  & Frontend / Backend containers
  & —
  & Ports 3000 (Next.js) and 3001 (Express); isolated service networks.
  & \ready
  & — \\

Orchestration (target)
  & Kubernetes
  & —
  & Production orchestration, scaling, rollouts.
  & \ready
  & AWS ECS/Fargate; Nomad; Fly.io; Render/Heroku (managed PaaS). \\

Ingress / Load Balancer
  & Nginx
  & —
  & Reverse proxy, TLS termination, basic routing.
  & \ready
  & Traefik; HAProxy; cloud LBs (ALB/GLB). \\

Network \& Governance
  & TLS 1.2+; RBAC/ABAC; tenant isolation
  & —
  & Encrypted transport, least-privilege, multi-tenant boundaries.
  & \ready
  & mTLS (Istio/Linkerd); OPA/Gatekeeper; Zero-Trust patterns. \\

Secrets Management
  & dotenv (env vars)
  & 16.4.5
  & Environment-based secrets/config for prototype.
  & \cond
  & HashiCorp Vault; SOPS+KMS; cloud secret managers (AWS/GCP/Azure). \\

Monitoring \& Logging
  & Health endpoints; Morgan; console logs
  & —
  & Liveness/readiness checks; HTTP and app logs.
  & \cond
  & OpenTelemetry; Prometheus+Grafana; ELK/OpenSearch; Sentry (app errors). \\

Tracing
  & —
  & —
  & Not implemented in prototype.
  & \cond
  & Jaeger or Tempo (via OpenTelemetry); Honeycomb (managed). \\

AI / ML Integrations
  & Hugging Face API; OpenAI API; custom services
  & —
  & Model access (pretrained); LLM playground; anomaly detection, preprocessing, security scanning, compliance auditor.
  & \ready
  & Inference gateways (vLLM/TGI/Ollama); Bedrock/Azure OpenAI; modal.run (serverless GPU). \\

External Data Sources
  & Avoindata.fi; Statistics Finland; THL Sotkanet; Findata; FSD; EU ODP; ELIXIR; CSC
  & —
  & National/EU datasets and HPC resources integrated where applicable.
  & \ready
  & Country-specific portals in new regions; commercial providers (e.g., Refinitiv). \\

Delivery \& CI/CD
  & Next.js build; multi-stage Docker images
  & —
  & Optimised prod images; reproducible builds.
  & \ready
  & Turborepo/Nx; GitHub Actions/GitLab CI; Argo CD/Flux (GitOps). \\

Edge / CDN
  & —
  & —
  & Not used in prototype.
  & \cond
  & Cloudflare/Akamai/Fastly; Vercel/Netlify Edge for SSR/ISR \& caching. \\

\end{longtable}}

\chapter{Governance and Policy Framework}
\label{sec:governance}

\fcolorbox{brand}{panel}{%
  \begin{minipage}{0.97\textwidth}
  \vspace{0.6em}
  \textbf{Highlights}
  \begin{itemize}
    \item Purpose and scope of the governance and policy framework
    \item Neutral, transparent, and accountable operating model
    \item Clearly defined roles, responsibilities, and approval processes
    \item Fair access and resource governance across participants
    \item Comprehensive data governance aligned with GDPR and the EU AI Act
    \item Balanced IP, licensing, and collaboration policies
    \item Built-in compliance, assurance, and reporting mechanisms
    \item Security baseline and incident–response procedures
    \item Change control, versioning, and policy–technology linkage
  \end{itemize}
  \vspace{0.4em}
  \end{minipage}
}

\section{Purpose and Scope}
The purpose of this section is to develop a governance and policy framework that will guide
how the AI Sandbox is managed and used in practice. The framework is not only meant to
set rules but also to create the conditions for trust, transparency, and neutrality across
all participating organisations. It describes how responsibilities are shared, how
access to resources is granted, and how decisions are made so that the platform remains
fair and accountable. The scope of the framework covers all types of participants, including universities, companies, public organisations, and individual researchers. It ensures that their activities within the sandbox are supported by clear policies for data handling, resource allocation, and collaboration. In particular, the framework integrates requirements from the GDPR and the EU AI Act (risk-based obligations), while also addressing intellectual property, data protection, and practical compliance needs \cite{gdpr,EUAIAct2024}.

\subsection{Operating Model and Neutrality}
The proposed operating model of the AI Sandbox is designed to guarantee neutrality, transparency, and long-term sustainability. In this section we outline how the sandbox 
will be governed, who will oversee its operations, and how conflicts and decisions will be managed. The model is intentionally built to avoid vendor lock-in and to ensure that 
all participating organisations can rely on a fair and transparent platform.

\begin{itemize}
\item \textbf{Neutral Operator:} We propose that the sandbox should be operated as a neutral entity, for example a consortium, foundation, or university-operated service with a clear neutrality mandate. The operator should avoid vendor lock-in and ensure openness of policies, interfaces, and technical components. The operator handles day-to-day administration and publishes policy updates and system changes in a transparent and accessible manner.
\item \textbf{Steering Group:} We propose the establishment of a multi-stakeholder Steering Group composed of representatives from academia, SMEs, larger industry, and the public sector. Its role is to provide strategic oversight, approve major policy changes, and arbitrate disputes. Meetings should be held quarterly, with consensus as the primary decision-making approach and majority voting as a fallback.
 \item \textbf{Conflict of Interest:} To ensure fair decision-making, all Steering Group members should be required to declare relevant financial, institutional, or personal interests. In cases where conflicts arise, members must recuse themselves from related deliberations or decisions \cite{oecd-coi}.
 This process reduces the risk of biased outcomes and strengthens stakeholder trust in the governance structure.
 \item \textbf{Accountability \& Reporting:} We propose that both the operator and the Steering Group issue regular reports documenting decisions, usage statistics, financial expenditures, and compliance audits. These reports should be shared with all stakeholders to maintain transparency, demonstrate fairness in resource allocation, and ensure that the sandbox evolves in line with community needs.
\end{itemize}
\begin{figure}[t]
\centering
\includegraphics[width=0.5\textwidth]{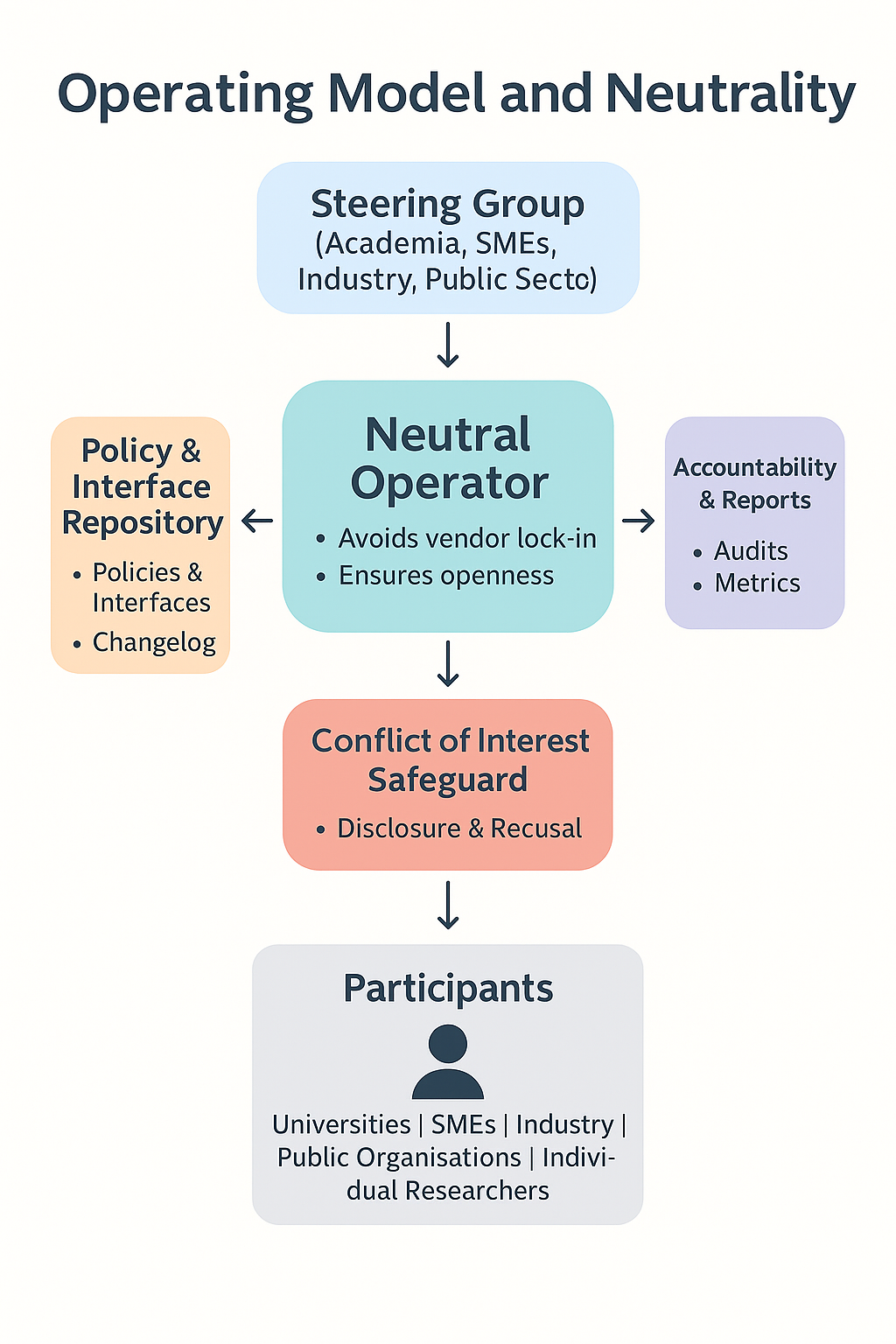}
\caption [Proposed Operating Model]{Proposed Operating Model and Neutrality Structure for the AI Sandbox, with oversight by the Steering Group, day-to-day administration by a Neutral Operator, and safeguards through policy publication, accountability reporting, and conflict-of-interest management (AI-assisted illustration)}
\label{fig:operatingmodel}
\end{figure}

\subsection{Roles, Responsibilities, and RACI}
To operationalise the proposed governance model, we also propose to define clear roles and
responsibilities for all participants in the sandbox. While the previous subsection
outlined who provides strategic oversight, this section addresses how responsibilities are
distributed in day-to-day operations. We adopt a RACI framework \footnote{https://project-management.com/understanding-responsibility-assignment-matrix-raci-matrix/}(Responsible, Accountable,
Consulted, Informed) to make explicit who performs which actions, who provides oversight,
and who must be kept informed. This approach supports transparency, reduces overlaps, and
builds confidence among stakeholders \cite{raci-explainer}.

\subsubsection{User Catalogue}
We propose that the sandbox adopts a comprehensive role catalogue aligned with the extended
access control system demonstrated in the prototype. This ensures that academic,
corporate, individual, technical, and external actors are formally represented, each with
clear responsibilities and access rights. The proposed categories are:
\begin{itemize}
\item \textbf{System Roles:} Super administrator, research administrator, researcher, viewer.
\item \textbf{Institutional Roles:} University administrator, coordinator, faculty, researcher, student.
\item \textbf{Corporate Roles:} Corporate administrator, manager, researcher, analyst, intern.
\item \textbf{Individual Roles:} Independent researcher, consultant, postdoc, visiting scholar.
\item \textbf{Technical Roles:} Data scientist, machine learning engineer, AI researcher, security analyst.
\item \textbf{Platform Roles:} Platform moderator, support, auditor.
\item \textbf{External Roles:} Government official, regulatory officer, funding agency, industry partner.
\end{itemize}
The prototype access-control design supports these categories through role-based and attribute-based controls, and is intended to scale as new stakeholder needs emerge.

\begin{table}[H]
\centering
\scriptsize
\caption[Illustrative RACI]{Illustrative RACI for Key Governance Activities in the AI Sandbox. 
RACI notation: \textbf{R} = Responsible, \textbf{A} = Accountable, 
\textbf{C} = Consulted, \textbf{I} = Informed. When \textbf{A/R} appears together, 
it indicates that the same actor is both Accountable and Responsible for the activity.}
\begin{tabular}{p{4.1cm} p{2.3cm} p{2.3cm} p{2cm} p{2cm} p{2cm}}
\toprule
\textbf{Activity} & \textbf{Platform Operator} & \textbf{Org Admin (Univ/Corp)} & \textbf{Admin / PI} & \textbf{Compliance} & \textbf{Steering Group} \\
\midrule
User onboarding approval & A/R & R & C & C & I \\
Dataset ingestion approval & C & A/R & R & C & I \\
GPU quota increase & C & A/R & R & C & I \\
Project release/sign-off & C & A/R & R & C & I \\
Policy changes (major) & C & I & I & C & A \\
Incident response (security) & C & C & I & A/R & I \\
Compliance audit \& evidence & C & R & R & A & I \\
\bottomrule
\end{tabular}
\label{tab:raci}
\end{table}

\subsubsection{Approval Authorities}
We propose that the RACI framework is supported by structured approval authorities, which
balance efficiency with accountability. The model automates low-risk processes while
maintaining human oversight where sensitivity or risk is higher:
\begin{itemize}
\item \textbf{Onboarding:} Low-risk roles (e.g., students, viewers) may be approved
automatically, while sensitive roles (e.g., corporate administrators, external
stakeholders) require review by an Organisation Admin. Research administrators
may co-approve onboarding requests within their projects to ensure smooth
academic workflows.
\item \textbf{Datasets:} Organisation Admins are accountable for dataset ingestion,
supported by Security/Compliance experts where sensitive or regulated datasets
are involved. Principal Investigators remain responsible for ensuring scientific
justification of dataset use.
\item \textbf{Compute Quotas:} Requests for increased GPU/CPU allocations should be
reviewed by Organisation Admins, with the Platform Operator consulted to confirm
capacity and Security/Compliance consulted to ensure isolation and data
protection requirements are upheld.
\end{itemize}

\subsection{Access and Resource Governance}

To ensure that the AI Sandbox remains both flexible and sustainable, access to 
resources is governed through a model that balances fairness, scalability, and 
compliance. Rather than prescribing fixed hardware quotas that may quickly become 
outdated, the framework defines resource classes, allocation tiers, and service 
levels that can evolve with available capacity and user demand. This approach 
reflects practices in established research and cloud infrastructures \cite{csc-elixir,elixir-finland}.

\subsubsection{Project Types}
We propose that the sandbox supports both \emph{shared} and \emph{private/isolated} 
project environments. Shared projects enable collaborative experimentation, while 
isolated projects are reserved for sensitive or regulated data scenarios and include 
stricter logging, audit, and access controls. Project type is determined at the 
onboarding stage and can be adjusted through approval workflows.

\subsubsection{Resource Classes and Allocation Tiers}
Resources are allocated in classes, with quotas scaling by role and subscription 
tier. These include:
\begin{itemize}
\item \textbf{Compute:} Access to CPU and GPU resources, with allocations ranging 
from individual fair-share use to organisation-level pools. GPU resources support 
both training and inference of large language models.
\item \textbf{Storage:} Tiered storage combining high-performance file systems 
(for active experimentation), object storage (for datasets and models), 
and archival storage (for long-term preservation). Redundancy and encryption 
are mandatory across tiers.
\item \textbf{Cloud/Hybrid:} Integration with Kubernetes-based environments 
(AWS, Azure, GCP, or national providers) for containerised AI workloads. 
Enables distributed training, hybrid deployment, and portability of workflows. Specific limits (e.g., GPU hours, storage quotas) are defined in the operational SLA and can be revised as infrastructure evolves. All allocations are fully auditable.
\end{itemize}

\begin{table}[H]
\centering
\scriptsize
\caption{Indicative Allocation Tiers and Service Levels by Role}
\begin{tabular}{p{3cm} p{3cm} p{3cm} p{3cm} p{3cm}}
\toprule
\textbf{Resource / SLA} & \textbf{Researcher} & \textbf{Research Admin / PI} & \textbf{Org Admin (Univ/Corp)} & \textbf{Notes / SLA Target} \\
\midrule
Compute (CPU/GPU) & Elastic access, fair-share & Higher-tier access, multi-GPU jobs & Org-wide pools with delegated quotas & Standard approvals $\leq$ 48h \\
Storage & Tiered (working + archive) & Extended tier with project datasets & Org-level allocation with flexible scaling & Data redundancy + encryption \\
Concurrent jobs & Limited concurrent jobs & Expanded concurrent capacity & Org-wide concurrent allocation & Fair-share scheduling applied \\
Cloud / Hybrid Access & Pre-configured templates & Custom container workflows & Org-level integration with external cloud credits & Elastic scaling where available \\
Incident response & --- & --- & --- & Critical security incidents handled $\leq$ 24h \\
Audit / Reporting & User-level usage logs & Project-level reports & Org-wide usage and compliance dashboards & All logs auditable by operator \\
\bottomrule
\end{tabular}
\label{tab:quotas-sla}
\end{table}

\paragraph{Usage Policies}
\begin{itemize}
\item An acceptable-use policy prohibits misuse of resources, re-identification 
attempts, or circumvention of security and compliance controls.
\item All approvals, usage, and cost attributions are logged and reportable per 
project, providing a transparent audit trail for accountability.
\item Resource allocations are subject to periodic review to ensure fair access 
across users and alignment with community priorities.
\end{itemize}

\subsection{Data Governance}
\begin{figure}[H]
\centering
\includegraphics[width=1\textwidth]{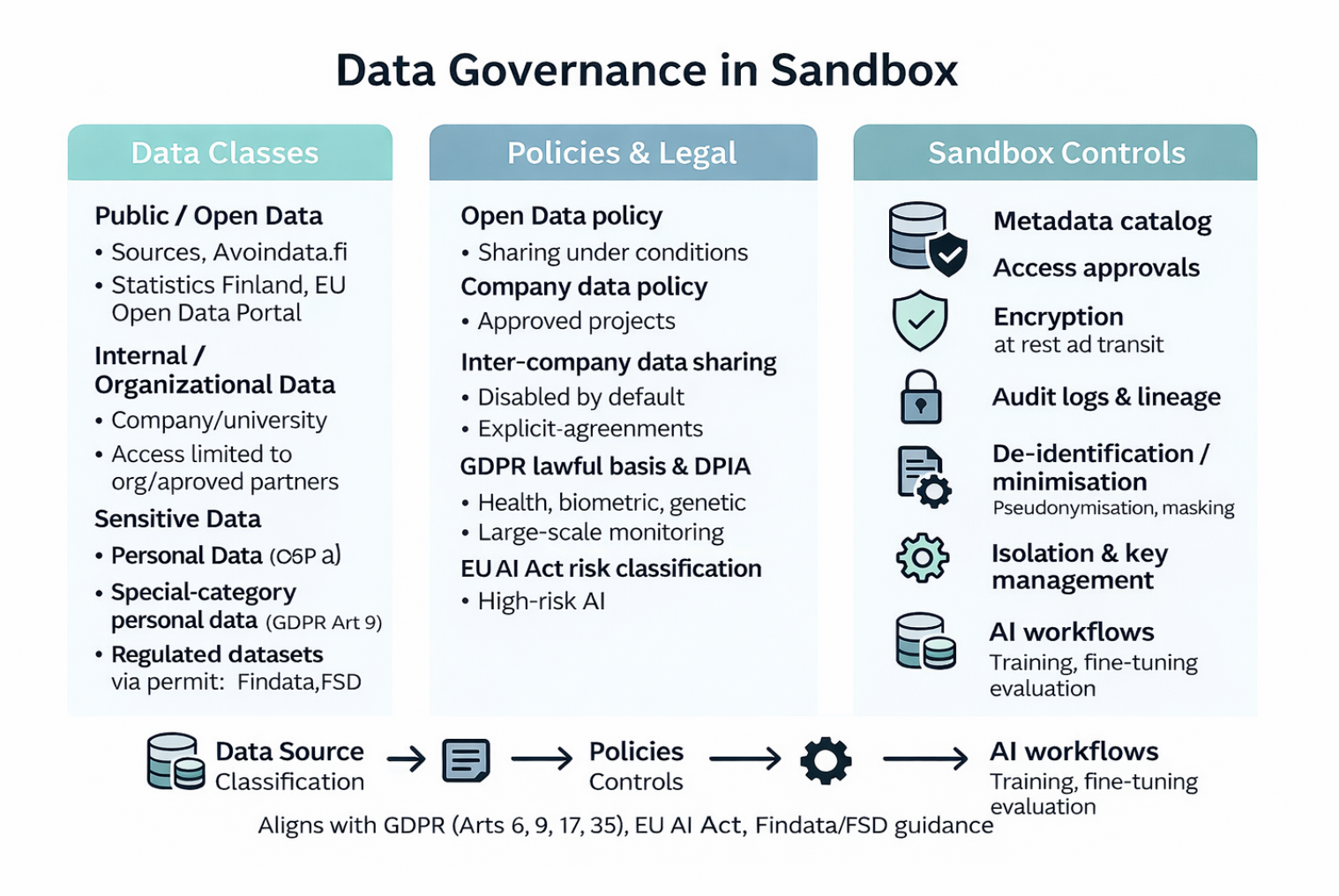}
\caption[Data governance framework]{Data governance framework in the AI Sandbox: classes and residency, 
policies for open/company/inter-company data, lawful bases and DPIA, and platform 
controls (metadata catalogue, access approvals, encryption, audit logs, lineage, 
and secure retention). (AI-assisted illustration)}
\label{fig:datagov}
\end{figure}

Building on the access and resource controls, we propose a comprehensive data governance 
framework that makes lawful basis, residency, approvals, and lifecycle operations explicit 
for every dataset ingested into the AI Sandbox. The framework is aligned with established 
regulatory and industry standards, including the \textbf{GDPR} (Articles 6, 9, 26, 28, 32, 35, 
and 89) and the \textbf{EU AI Act} (Regulation (EU) 2024/1689, with phased application), together with \textbf{ISO/IEC 27001} and \textbf{27701} for 
information security and privacy management, and \textbf{ISO/IEC 23894:2023} for AI risk 
management \cite{gdpr,EUAIAct2024,ISO23894}. It also draws on national guidance from \textbf{Findata}, \textbf{THL}, and the 
\textbf{Finnish Social Science Data Archive (FSD)} for handling sensitive health and social 
data. Figure~\ref{fig:datagov} illustrates the overall control stack, while the detailed 
policy below operationalises it.

\subsubsection{Data Classes}
For governance purposes, all datasets within the AI Sandbox are organised into 
standardised data classes that build directly on the sources identified in the 
data management section. This classification provides a clear link between the 
origin of the data, the associated legal and contractual requirements, and the 
technical controls applied in the sandbox. It ensures that policies are applied 
consistently across heterogeneous sources and that compliance can be enforced 
through transparent rules.

\begin{itemize}
\item \textbf{Public / Open Data:} Includes datasets from national and European 
open data portals such as \emph{Avoindata.fi}, \emph{Statistics Finland}, and 
the \emph{EU Open Data Portal}. These datasets are subject to open licences 
(e.g., CC BY, CC0) that define reuse conditions. They may be broadly used 
for experimentation and benchmarking, with licence terms recorded in the 
sandbox metadata catalogue \cite{ccby,cc0}. 
\item \textbf{Internal / Organisational Data:} Refers to datasets contributed by 
participating organisations (companies, universities, or public bodies). 
Access is restricted to the owning organisation and named collaborators. 
A data processing agreement (DPA) or equivalent intra-group policy must 
be in place, and default residency is within the EU/EEA.
\item \textbf{Sensitive Data:} Covers data subject to special safeguards, such as 
health, social, or personal data obtained from \emph{Findata}, the 
\emph{Finnish Social Science Data Archive (FSD)}, the \emph{Finnish Institute 
for Health and Welfare (THL)}, or biomedical datasets accessed via 
\emph{ELIXIR/CSC}. These datasets can only be processed under the authority’s 
permit or licensing conditions. Secure, isolated workspaces are mandatory, 
and all exports are subject to disclosure control and review \cite{elixir-finland}.
\end{itemize}

\subsubsection{Roles and Responsibilities (GDPR alignment)}
Clear assignment of roles and responsibilities is central to accountability under GDPR and the EU AI Act. We propose the following allocation, ensuring that controllers, processors, and compliance officers act in line with their legal duties \cite{gdpr,EUAIAct2024}.
\begin{itemize}
\item \textbf{Controller}: By default, the organisation that contributes a dataset 
(company, university, or public body). For joint projects, the parties may act 
as joint controllers under GDPR Article~26, with a written allocation of responsibilities \cite{gdpr}.
\item \textbf{Processor}: The Sandbox Operator acts as a processor for controller datasets, providing the technical and organisational measures required by GDPR Article~28 \cite{gdpr}.
\item \textbf{DPO \& Security/Compliance}: Review data processing agreements, sign off 
on DPIA requirements under Article~35, and audit compliance evidence \cite{gdpr,edpb-dpia}.
\end{itemize}

\subsubsection{Policies by Source Type}
Different categories of data require tailored governance rules. The following policies cover open/public datasets, internal company or university data, sensitive datasets, and inter-company data sharing.
\begin{itemize}
\item \textbf{Open/Public data} (Avoindata.fi, Statistics Finland, EU Open Data 
Portal): May be used for research and method evaluation subject to the original 
licence (e.g., CC BY/CC0). Licences are recorded in the metadata catalogue; 
redistribution terms and attribution requirements are enforced in outputs.
\item \textbf{Internal/Organisational data} (company or university): Access is 
restricted to the owner organisation and named collaborators. A Data Processing 
Agreement (DPA) or intra-group policy is required; default residency is EU/EEA.
\item \textbf{Sensitive data} (Findata, FSD/THL, ELIXIR context): Processing follows the originating authority’s permit or conditions. Secure, isolated workspaces 
are mandatory; exports require documented disclosure control and permit-compliant review before release.
\item \textbf{Inter-company sharing}: Disabled by default. Allowed only via a signed Data Sharing Agreement (DSA), per-dataset access lists, and a recorded lawful basis (e.g., contract, consent, public task). High-risk categories trigger a Data Protection Impact Assessment (DPIA) \cite{gdpr,edpb-dpia}.
\end{itemize}

\subsubsection{Classification and Residency}
To ensure lawful and consistent handling, all datasets are classified and assigned a 
default residency. This classification provides the foundation for storage, processing, 
and cross-border data management rules.
\begin{itemize}
\item \textbf{Classes}: \emph{Public/Open} (licence-governed, non-personal), 
\emph{Internal/Organisational} (non-public, may include personal data), 
\emph{Sensitive} (e.g., health, social, minors, special categories of personal 
data as per GDPR Article~9).
\item \textbf{Residency}: Sensitive data are stored and processed in the \textbf{EU/EEA} by
default; internal data follow contractual terms; public/open data may be hosted more broadly
if licence permits. Residency is declared in metadata and enforced by policy controls \cite{gdpr}.
\end{itemize}

\subsubsection{Lawful Bases, Agreements, and DPIA}
Every dataset must have a documented lawful basis and, where required, supporting 
agreements or permits. DPIA requirements are triggered in high-risk contexts and must 
be completed before project activation.
\begin{itemize}
\item Each dataset is registered with its lawful basis (e.g., consent, public interest/public task, legitimate interest, contract) and linked agreements (DPA/DSA/permit) \cite{gdpr}. 
\item \textbf{DPIA triggers}: health/biometric or other special-category data; 
large-scale monitoring; vulnerable subjects; cross-organisation sharing; high-risk AI use. DPIA outcomes, including mitigations and human-oversight measures, are recorded before project activation \cite{gdpr,edpb-dpia}.
\end{itemize}

\subsubsection{Technical \& Operational Controls}
To make policies enforceable in practice, the sandbox implements technical and 
operational controls for dataset registration, access approvals, encryption, isolation, 
and anonymisation.
\begin{itemize}
\item \textbf{Dataset registration \& metadata}: Mandatory metadata catalogue 
(source, licence/permit, controller/processor roles, lawful basis, residency, 
retention, contact, lineage identifiers).
\item \textbf{Access approvals}: Role-based approvals per dataset; time-bound access 
grants; per-project access lists; automatic expiry and re-certification.
\item \textbf{Isolation \& least privilege}: Project namespaces, network policies, and 
storage isolation; secrets management for credentials; audit-ready logs capturing 
who, what, when, and why.
\item \textbf{Protection}: Encryption at rest and in transit (AES-256, TLS 1.2+); key 
management using managed KMS/HSM; integrity checksums on artefacts.
\item \textbf{De-identification}: Pseudonymisation/anonymisation tooling 
(e.g., k-anonymity, $\ell$-diversity, t-closeness where applicable) with 
residual-risk notes attached to the dataset record.
\end{itemize}

\subsubsection{Retention, Deletion, and Lineage}
Retention, deletion, and lineage controls ensure that data is not held longer than 
necessary and that its lifecycle is auditable end-to-end. This supports both compliance 
and reproducibility of experiments.
\begin{itemize}
\item \textbf{Retention}: Defined per project and source terms (e.g., permit end date). 
Tiered storage may be used for cost control, subject to residency limits.
\item \textbf{Deletion \& exit}: Cryptographic erasure or secure wipe at project 
closure, with a verifiable \emph{certificate of destruction}. Model artefacts 
derived from sensitive data are reviewed for disclosure risk before retention 
or release.
\item \textbf{Lineage}: End-to-end lineage maintained (source $\rightarrow$ 
preprocessing $\rightarrow$ training/evaluation $\rightarrow$ outputs) to 
support reproducibility, audits, and GDPR challenge/appeal rights.
\end{itemize}

\subsection{IP and Licensing Policy}
Intellectual property (IP) and licensing in the AI Sandbox must balance openness, 
innovation, and legal certainty for all participants. The proposed policy builds on established practices in Finnish and EU collaborative R\&D projects and aligns with European and national legal frameworks \cite{tradesecrets,ec-ip-action,he-mga}, including the EU Directive on the protection of trade secrets\footnote{Directive (EU) 2016/943 on the protection of undisclosed know-how and business information (trade secrets).}, the EU Intellectual Property Action Plan\footnote{European Commission, ``Making the most of the EU’s innovative potential: An intellectual property action plan to support the EU’s recovery and resilience,'' COM(2020) 760 final, November 2020.}, Finnish copyright and data rights 
legislation, and model agreements such as those used in Horizon Europe\footnote{European 
Commission, Horizon Europe Model Grant Agreement, 2021.}. The framework ensures that 
contributors retain their rights, that new results are clearly allocated, and that 
licensing remains transparent, auditable, and enforceable.

\begin{figure}
\centering
\includegraphics[width=0.8\textwidth]{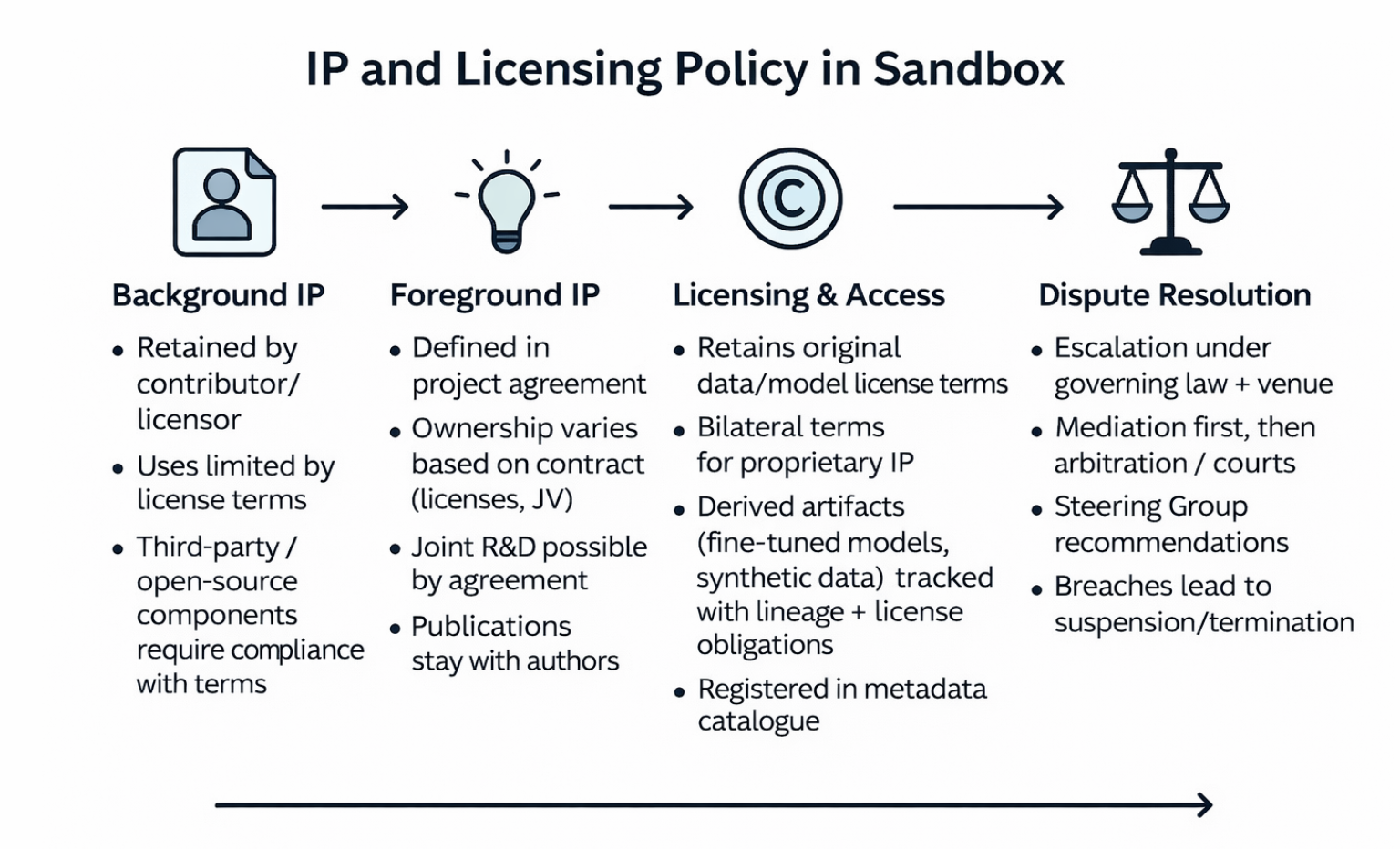}
\caption[Intellectual Property (IP) and Licensing Policy]{Intellectual Property (IP) and Licensing Policy in the AI Sandbox, showing allocation of background and foreground IP, licensing of tools and deliverables, and enforcement mechanisms aligned with EU and Finnish legal frameworks. (AI-assisted illustration)}
\label{fig:placeholder}
\end{figure}

\subsubsection{Background IP}
Background IP refers to the rights that participants bring into the sandbox, such as 
datasets, algorithms, or software. The policy preserves ownership while allowing 
limited use within projects.
\begin{itemize}
\item Background IP remains the property of the original licensor or contributor. 
\item Usage rights are granted only to the extent required for the project, under 
licence terms or data processing agreements (DPAs). 
\item Confidential or trade-secret materials are protected by law and may not be 
redistributed without explicit permission. 
\end{itemize}

\subsubsection{Foreground IP}
Foreground IP is generated in sandbox projects, including trained models, source 
code, and derived datasets. Ownership follows transparent and predictable rules.
\begin{itemize}
\item Foreground IP is owned by the project sponsor by default, unless otherwise 
specified in writing. 
\item Joint ownership may be agreed for collaborative R\&D, with clear provisions 
on exploitation, licensing, and revenue sharing. 
\item Academic publications remain with the authors but must acknowledge the sandbox 
infrastructure. 
\end{itemize}

\subsubsection{Tools, Components, and Open Source}
Core sandbox infrastructure is treated as a shared public good to avoid vendor lock-in 
and ensure long-term neutrality.
\begin{itemize}
\item Core sandbox tools (platform services, APIs, orchestration scripts) should be 
released under permissive licences such as Apache-2.0 or MIT. 
\item Project-specific deliverables (e.g., fine-tuned LLMs) remain proprietary unless 
explicitly released under open-source or open-data terms. 
\item All third-party dependencies are recorded in a Software Bill of Materials (SBOM) 
to ensure licence compliance and vulnerability management. 
\end{itemize}

\subsubsection{Licensing and Access Conditions}
Licensing conditions define how sandbox outputs may be reused across research, 
industry, and public contexts.
\begin{itemize}
\item Open/public data reuse follows the original licence terms (e.g., CC BY/CC0) \cite{ccby,cc0}. 
\item Proprietary results may be licensed bilaterally under sandbox agreements, with 
different terms for research use versus commercial exploitation. 
\item All licences are registered in the sandbox metadata catalogue and validated as 
part of the approval workflow. 
\end{itemize}

\subsubsection{Dispute Resolution and Enforcement}
Clear mechanisms are needed to build trust and address conflicts between participants.
\begin{itemize}
\item Disputes are governed by Finnish law, with mediation preferred and arbitration 
or court proceedings as fallback mechanisms. 
\item The multi-stakeholder Steering Group may issue non-binding recommendations to 
preserve neutrality and avoid escalation. 
\item Violations of licence terms may trigger audits, suspension of access, or 
termination of project participation. 
\end{itemize}

\subsection{Compliance and Assurance}
To ensure that the AI Sandbox operates within a legally robust and trustworthy
framework, we propose a layered compliance and assurance model aligned with the
requirements of the EU AI Act, the GDPR, Finnish data protection legislation, and
established international standards such as ISO/IEC 27001 (information security
management) and ISO/IEC 23894:2023 (AI risk management). This framework is designed to provide transparency, verifiable accountability, and confidence for all participating stakeholders \cite{EUAIAct2024,gdpr,iso27001}.

\subsubsection{EU AI Act Alignment}
The sandbox adopts the risk-based approach of the EU AI Act to classify AI systems and apply safeguards proportionate to their risk profile \cite{EUAIAct2024}.

\begin{itemize}
\item \textbf{Risk classification:} All AI projects in the sandbox are registered
with a declared risk category (minimal, limited, or high-risk), based on the
intended use case, data sensitivity, and potential societal impact.
\item \textbf{High-risk controls:} Projects flagged as high-risk (e.g., involving
biometric data, employment screening, or health-related decision support) must
include human oversight checkpoints, expanded logging, robustness testing, and
explainability measures before deployment or external release.
\item \textbf{Compliance evidence:} Documentation of conformity-related artefacts and DPIAs
(Data Protection Impact Assessments) are recorded in the sandbox metadata
catalogue and reviewed by security/compliance officers prior to project
activation.
\end{itemize}

\subsubsection{Evidence and Auditability}
Trust in the sandbox depends on transparent and verifiable processes for monitoring
compliance and demonstrating accountability to regulators and stakeholders.
\begin{itemize}
\item \textbf{Immutable logging:} All key events, including approvals, dataset
access, training runs, model evaluation, and release decisions, are logged
in an immutable and tamper-evident audit trail.
\item \textbf{Internal audits:} The sandbox operator will perform periodic internal
audits of compliance, security, and data protection practices at least twice
per year, with findings reported to the Steering Group.
\item \textbf{External assurance:} Independent third-party audits or certifications
(e.g., ISO/IEC 27001 or national security audits) may be commissioned by the
Steering Group to validate compliance and neutrality.
\item \textbf{Corrective actions:} Non-conformities identified in audits trigger
remediation plans with defined timelines, monitored by the compliance function
and reported back to stakeholders.
\end{itemize}

\begin{figure}
\centering
\includegraphics[width=0.9\linewidth]{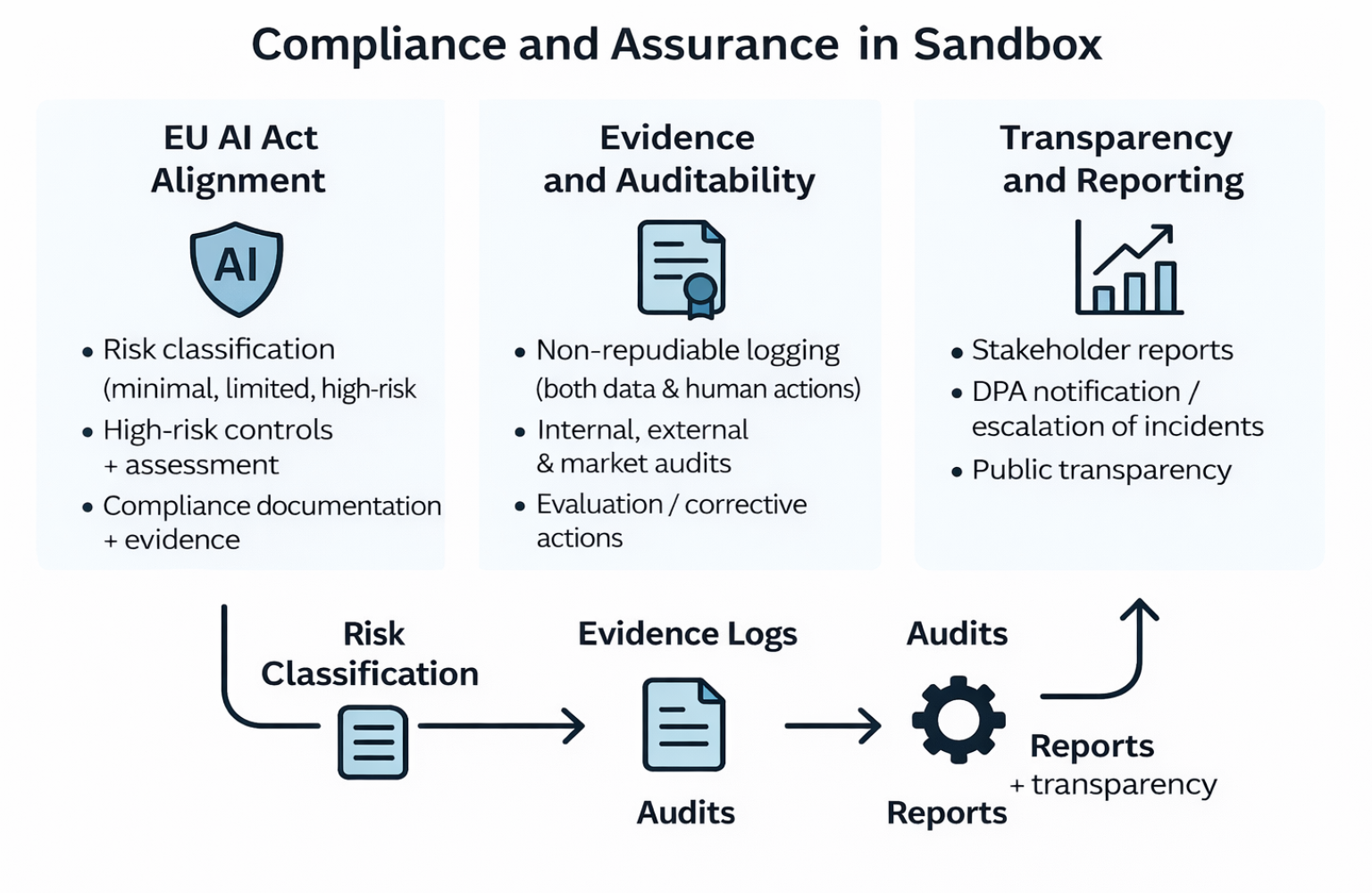}
\caption[Compliance and assurance framework]{Compliance and assurance framework in the AI Sandbox, showing alignment with the EU AI Act and GDPR through risk classification, human oversight, audit trails, and layered internal/external audits to support accountability and trust (AI-assisted illustration).}
\label{fig:Compliance}
\end{figure}

\subsubsection{Transparency and Reporting}
Regular reporting strengthens trust and ensures that compliance activities remain
visible to all participants.
\begin{itemize}
\item \textbf{Stakeholder reports:} Quarterly compliance summaries, including risk
classifications, audit outcomes, and corrective actions, are circulated to
participating organisations.
\item \textbf{Regulatory liaison:} Where applicable, reports are aligned with the
obligations under Finnish Data Protection Authority guidance and relevant EU
supervisory bodies.
\item \textbf{Public transparency:} Aggregated and anonymised reports (e.g., number
of projects by risk class, compliance audit pass rates) may be published on the
sandbox portal to demonstrate accountability to the wider ecosystem.
\end{itemize}

\subsection{Security Baseline and Incident Response}
Building on the operating model, RACI, and data governance controls, we propose the following security and incident–response baseline for the AI Sandbox. The baseline is designed to be vendor-neutral and auditable, and aligns with GDPR, the EU AI Act, and widely adopted security standards and guidance \cite{gdpr,EUAIAct2024,iso27001,NIS2,nist80061}. Where possible, the measures reference widely adopted norms (ISO/IEC 27001/27017/27018/27701; CIS Benchmarks; ENISA and NIST SP 800-61 guidance).

\subsubsection{Identity and Access Management (IAM)}
We will operate a least-privilege, role/attribute-based model consistent with the role catalogue and RACI.
\begin{itemize}
\item \textbf{Federated SSO \& MFA:} SAML/OIDC federation with home institutions and companies; MFA enforced for administrators, dataset uploaders, approvers, and any user with elevated privileges \cite{oidc-core,saml2-core}.
\item \textbf{Separation of duties:} Approval of onboarding, dataset ingestion, and quota changes requires distinct roles; no single user may both request and approve.
\item \textbf{Scoped credentials:} Short-lived, scoped API tokens; service accounts with minimal scopes; quarterly access recertification; automatic expiry and re-approval for time-bound access.
\item \textbf{Tenant isolation:} Organisation and project workspaces are isolated (RBAC/ABAC, namespace and network policies) with default-deny access between tenants.
\end{itemize}

\subsubsection{Data Security and Isolation}
Controls protect data in use, in transit, and at rest, and respect declared residency.
\begin{itemize}
\item \textbf{Encryption:} TLS~1.2+ for all connections; AES-256 (or provider equivalent) at rest. Keys handled by a managed KMS/HSM with EU/EEA residency and role-based key usage.
\item \textbf{Network controls:} Private subnets, security groups, VPC peering policies, private endpoints to storage; egress controls and DLP rules for sensitive projects; workload identity for intra-cluster access.
\item \textbf{Secrets management:} Central secrets manager; no secrets in code or images; automatic rotation for database/cloud keys.
\item \textbf{Workspace isolation:} Sensitive projects run in dedicated, hardened clusters/nodes with policy as code (Pod/Node Security, image allow-lists, resource quotas); explicit export gates for artefacts leaving an isolated workspace.
\item \textbf{Backups \& recovery:} Encrypted backups with EU/EEA residency; routine restore testing; defined RPO/RTO targets (see Operational Metrics).
\end{itemize}

\subsubsection{Assurance: Vulnerability \& Supply-Chain Management}
\begin{itemize}
\item \textbf{Vulnerability management:} Monthly authenticated scans; critical findings remediated within agreed SLOs (e.g., 7~days for critical, 30~days for high). Annual third-party penetration testing.
\item \textbf{SBOM \& SCA:} Software Bill of Materials for all platform components; continuous software-composition analysis for licence and CVE issues; container images built from minimal CIS-hardened bases, signed and verified (e.g., Sigstore).
\item \textbf{Change control:} Infrastructure-as-Code with peer review; staged rollouts; automated policy checks (OPA/Conftest) before deploy.
\end{itemize}

\subsubsection{Incident Response Playbook}
Our process follows ENISA and NIST~SP~800-61 phases and meets GDPR breach-notification timelines \cite{NIS2,nist80061,gdpr}.
\begin{itemize}
\item \textbf{Detect \& triage:} Centralised, immutable logs (WORM) and SIEM correlation for auth events, data access, training jobs, and model releases. Severity (S1--S4) classification with 24/7 on-call rotation for S1.
\item \textbf{Contain \& eradicate:} Isolation at tenant/project level; token/key revocation; image rollback; malicious artefact quarantining; temporary export blocks for affected datasets/models.
\item \textbf{Recover:} Verified restore from clean backups; integrity checksums for artefacts; stakeholder communications and service SLO recovery.
\item \textbf{Notify:} If personal data are implicated, the Controller is supported to assess notification to the authority within 72\,h (GDPR~Art.~33) and, where necessary, to data subjects (Art.~34) \cite{gdpr}.
\item \textbf{Learn:} Post-incident review within 10 working days; corrective actions tracked; semi-annual tabletop exercises for data breach and model-misuse scenarios.
\end{itemize}

\subsubsection{Addressing Cross-Company Collaboration Concerns}
The following map shows how the baseline mitigates the most frequent concerns when multiple companies collaborate in a shared LLM sandbox:
\begin{itemize}
\item \textbf{(1) Data security \& confidentiality:} Tenant isolation; RBAC/ABAC; encryption in transit/at rest with EU/EEA key residency; export gates and DLP for sensitive workspaces; dataset-level access lists tied to lawful basis and approvals.
\item \textbf{(2) IP ownership:} Background/foreground IP and licensing handled per Section~\emph{IP and Licensing Policy}; artefact provenance (dataset $\rightarrow$ training run $\rightarrow$ model) recorded to support ownership and attribution.
\item \textbf{(3) Model governance \& accountability:} Model registry with versioning; signed releases; prompt/dataset/job logging with privacy-preserving redaction; explainability notes for high-risk features; human-oversight checkpoints as required by the EU AI Act.
\item \textbf{(4) Interoperability \& access control:} Identity federation (SAML/OIDC), SCIM provisioning, standard APIs; organisation boundaries ensure partners do not require broad access to each other's systems.
\item \textbf{(5) Compliance \& legal risk:} DPIA triggers evaluated during intake; DPAs/DSAs linked to datasets; audit-ready evidence (approvals, lineage, logs); lawful-basis and residency enforced by policy as code.
\item \textbf{Cross-cutting: cost control:} Per-project budgets and alerts; fair-share scheduling; idle shutdown; reserved vs.\ on-demand GPU policies.
\item \textbf{Cross-cutting: data hygiene:} Intake QA checks (schema, PII scanners); required metadata completeness; prohibited-content filters for training corpora; dataset quality metrics tracked in the catalogue.
\end{itemize}

\subsubsection{Operational Metrics and SLAs}
To keep the baseline measurable, we propose the following initial targets (refined with the Steering Group during pilot):
\begin{itemize}
\item \textbf{Security events:} Mean time to detect (MTTD) S1 $\leq$ 4\,h; mean time to contain (MTTC) S1 $\leq$ 24\,h; incident report and lessons learned published within 10 business days.
\item \textbf{Patching:} Critical CVEs remediated within 7\,days; high within 30\,days.
\item \textbf{Backups:} RPO $\leq$ 24\,h; RTO $\leq$ 8\,h for core control plane; quarterly restore test.
\item \textbf{Access governance:} Quarterly access recertification; approval SLAs per Table~\ref{tab:quotas-sla}; automated expiry for time-bound grants.
\item \textbf{Logging retention:} Security logs $\geq$ 12\,months (immutable); dataset/model lineage retained for the life of the project plus contractual/archive period.
\end{itemize}

\subsubsection{Compliance Mapping (overview)}
\begin{itemize}
\item \textbf{GDPR:} Security of processing (Art.~32), breach notification (Arts.~33--34), records of processing (Art.~30), and privacy by design/default (Art.~25) operationalised through IAM, isolation, encryption, logging, and incident playbooks. Controllers remain responsible for legal bases; the Operator provides processor-level TOMs and evidence.
\item \textbf{EU AI Act (high-level):} For higher-risk features, the sandbox enables risk management and logging, technical documentation, transparency to users, human oversight checkpoints, and robustness/cybersecurity controls; conformity activities remain the responsibility of the provider/user of the AI system.
\item \textbf{Finnish practice for sensitive data:} Where applicable, Findata/THL/FSD permit conditions are enforced via isolated workspaces, export review, and documented disclosure control.
\end{itemize}

\begin{table}[H]
\centering
\scriptsize
\caption{Summary of Security Baseline and Incident Response Measures}
\begin{tabular}{p{3.2cm} p{6.2cm} p{4.2cm}}
\toprule
\textbf{Domain} & \textbf{Key Measures} & \textbf{Regulatory / Standard Alignment} \\
\midrule
Identity and Access Management & 
Multi-factor authentication (MFA) for admins, dataset uploaders, and approvers; strong password and session policies; role-based access with least privilege & 
GDPR Art. 32 (security of processing), ISO/IEC 27001 (A.9), NIST IAM controls \\
\addlinespace[4pt]
Data Security & 
Encryption at rest and in transit (AES-256, TLS 1.2+); network isolation for sensitive projects; secrets management; audit-ready logging of data access & 
GDPR Art. 25 (data protection by design), EU AI Act (data handling), ISO/IEC 27018 (cloud PII protection) \\
\addlinespace[4pt]
Assurance and Monitoring & 
Monthly automated vulnerability scans; annual penetration testing; continuous monitoring dashboards; Software Bill of Materials (SBOM) for dependencies & 
ISO/IEC 27001 (A.12, A.18), EU NIS2 Directive (cybersecurity baseline), ENISA cloud security guidelines \\
\addlinespace[4pt]
Incident Response & 
Structured lifecycle: triage $\rightarrow$ containment $\rightarrow$ eradication $\rightarrow$ recovery $\rightarrow$ post-mortem review; breach notifications within 72h; communication with stakeholders & 
GDPR Art. 33 (breach notification), ISO/IEC 27035 (incident response), NIS2 (reporting duties) \\
\bottomrule
\end{tabular}
\label{tab:security_incident}
\end{table}

\subsection{Change Control, Versioning, and Exit Strategy}
The governance framework of the AI Sandbox must evolve in step with regulatory,
technological, and organisational changes. To maintain continuity, transparency, and
legal certainty, we propose the following mechanisms:
\begin{itemize}
\item \textbf{Policy review and versioning:} Governance documents are reviewed at
least annually, or earlier if triggered by regulatory changes, newly identified
risks, or stakeholder requests. Semantic versioning (major/minor/patch) is
applied, and all historic versions are archived for reference.
\item \textbf{Change management:} Formal change proposals are submitted with defined
consultation windows for stakeholder input. The Steering Group approves
material amendments following documented deliberation. Minor technical or
editorial updates may be delegated to the Sandbox Operator, with notification
to all stakeholders.
\item \textbf{Exit and off-boarding:} Participants who leave the sandbox receive
support for secure data export, model handover, and transfer of relevant
metadata. Accounts are closed in a controlled process, including deletion or
cryptographic erasure of remaining data, with a verifiable certificate of
destruction issued.
\end{itemize}

\subsection{Implementation Link: Policy \texorpdfstring{$\leftrightarrow$}{↔} Technology}
To be effective, governance rules must be enforceable within the technical fabric of
the sandbox. We therefore propose a continuous mapping of policy requirements to
concrete platform controls:
\begin{itemize}
\item \textbf{Access control:} Role-based access control (RBAC) and multi-level
approval workflows enforce the principles defined in the governance model.
\item \textbf{Data governance:} Storage and processing are classified by data type
(public, internal, sensitive), with residency and retention controls enforced
automatically.
\item \textbf{Evidence capture:} Experiment tracking and system logs provide
immutable records of approvals, dataset use, model training, and releases,
supporting audits and regulatory compliance.
\item \textbf{Compliance dashboards:} Real-time dashboards summarise quotas,
approvals, data flows, and compliance indicators, allowing administrators and
auditors to monitor adherence to both sandbox policies and EU/Finnish
regulatory obligations.
\end{itemize}
This linkage is maintained as a \emph{living appendix} to the governance framework,
updated alongside each platform release, ensuring that changes in regulation or policy
translate directly into enforceable and auditable technical measures.

\section{Glossary of Abbreviations}
\begin{table}[H]
\centering
\scriptsize
\caption[Security and Compliance]{Glossary of Abbreviations and Definitions Used in the Governance Framework}
\begin{tabular}{p{3.2cm} p{11.5cm}}
\toprule
\textbf{Abbreviation} & \textbf{Definition / Reference} \\
\midrule
\textbf{GDPR} & General Data Protection Regulation (Regulation (EU) 2016/679). Core EU law on data protection and privacy. \\
\textbf{EU AI Act} & Artificial Intelligence Act, Regulation (EU) 2024/1689 of the European Parliament and of the Council of 13 June 2024 laying down harmonised rules on artificial intelligence (risk-based framework with phased application). \\
\textbf{DPA} & Data Processing Agreement, required by GDPR Art.~28 between data controllers and processors, defining responsibilities and safeguards. \\
\textbf{DSA} & Data Sharing Agreement, contractual mechanism to allow inter-organisational data access while defining lawful basis, security measures, and responsibilities. \\
\textbf{DPIA} & Data Protection Impact Assessment, a mandatory risk assessment for high-risk data processing under GDPR Art.~35. \\
\textbf{SBOM} & Software Bill of Materials, a formal record of components, dependencies, and licences in a software product (ISO/IEC 5962:2021). \\
\textbf{IAM} & Identity and Access Management: security processes and tools for role-based and attribute-based access, including authentication and authorisation. \\
\textbf{MFA} & Multi-Factor Authentication: security method requiring multiple credentials for access (e.g., password + hardware token). \\
\textbf{RPO} & Recovery Point Objective: maximum tolerable period in which data might be lost due to an incident. \\
\textbf{RTO} & Recovery Time Objective: target time to restore systems after a disruption. \\
\textbf{NIS2 Directive} & Directive (EU) 2022/2555 on measures for a high common level of cybersecurity across the Union (successor to the NIS Directive). \\
\textbf{ENISA} & European Union Agency for Cybersecurity. Provides guidelines, incident response recommendations, and security baselines. \\
\textbf{ISO/IEC 27001} & International standard for information security management systems (ISMS). Basis for many controls cited in the framework. \\
\textbf{ISO/IEC 27701} & Privacy extension to ISO/IEC 27001, providing requirements for data privacy management. \\
\textbf{ISO/IEC 23894:2023} & International standard for AI risk management, aligned with risk-management obligations discussed in the EU AI Act context. \\
\textbf{NIST SP 800-61} & US National Institute of Standards and Technology Special Publication 800-61: Computer Security Incident Handling Guide. Provides phases for incident response. \\
\bottomrule
\end{tabular}
\label{tab:abbreviations}
\end{table}

\chapter{AI Sandbox MVP}
\label{sec:implementation}

\section{Overview}

The current Minimum Viable Product (MVP) of the AI Sandbox demonstrates how secure, compliant, and collaborative AI experimentation can be achieved within a unified technical and governance framework. 
The system is deployed on the CSC Rahti cloud platform\footnote{\url{https://gptlab-frontend-gptlab-sandbox.2.rahtiapp.fi/collaborations/discovery}} and its source code, configuration scripts, and operational documentation are maintained in the GPT Lab repository\footnote{\url{https://github.com/GPT-Laboratory/GPT-Lab-Sandbox}}. 
A complementary Wiki provides guidance for deployment on institutional or private infrastructures, supporting reproducibility across research and industrial partners.

The MVP realises the core functions of the Sandbox: identity and access management, project governance, AI-service orchestration, hardware resource control, and collaboration hubs for universities and companies. 
The following sections illustrate the working prototype through representative screenshots and describe how the architecture can scale toward production.

\section{System Architecture and Key Components}

The deployed environment follows a modular microservices architecture designed for extensibility and security. 
Each service runs as an independent container within a Kubernetes-based platform on CSC Rahti, connected through secured REST APIs. 
The main components include:

\begin{itemize}
  \item \textbf{Access Gateway and Landing Interface:} Entry point with authentication and access controls. In the current prototype, access is handled through the platform’s account workflows and token-based sessions; institutional federation (for example via OpenID Connect) is treated as an integration path for production deployments.
  \item \textbf{Academic and Organisational Hubs:} Controlled workspaces for universities, companies, and joint projects with fine-grained access policies.
  \item \textbf{Dashboard and User Management:} Central administration for users, roles, and permissions across tenants.
  \item \textbf{Hardware and Resource Management:} Interfaces for GPU/CPU allocation, quota monitoring, and usage analytics.
  \item \textbf{AI Services and Model Evaluation:} Execution and benchmarking environment for LLMs and other AI models.
  \item \textbf{Data and Compliance Layer:} Enforcement of encryption, residency, and lawful-basis policies defined in the governance framework.
\end{itemize}

\section{User Interface Demonstration}

The following figures illustrate selected interfaces of the deployed Sandbox MVP. 
They reflect the role-based experience and show functional integration of security, usability, and collaboration.

\subsection{Landing Page and Access Gateway}

The landing interface provides authenticated access to the Sandbox and introduces available services for academic and industrial participants.

\begin{figure}[H]
  \centering
  \includegraphics[width=0.7\textwidth]{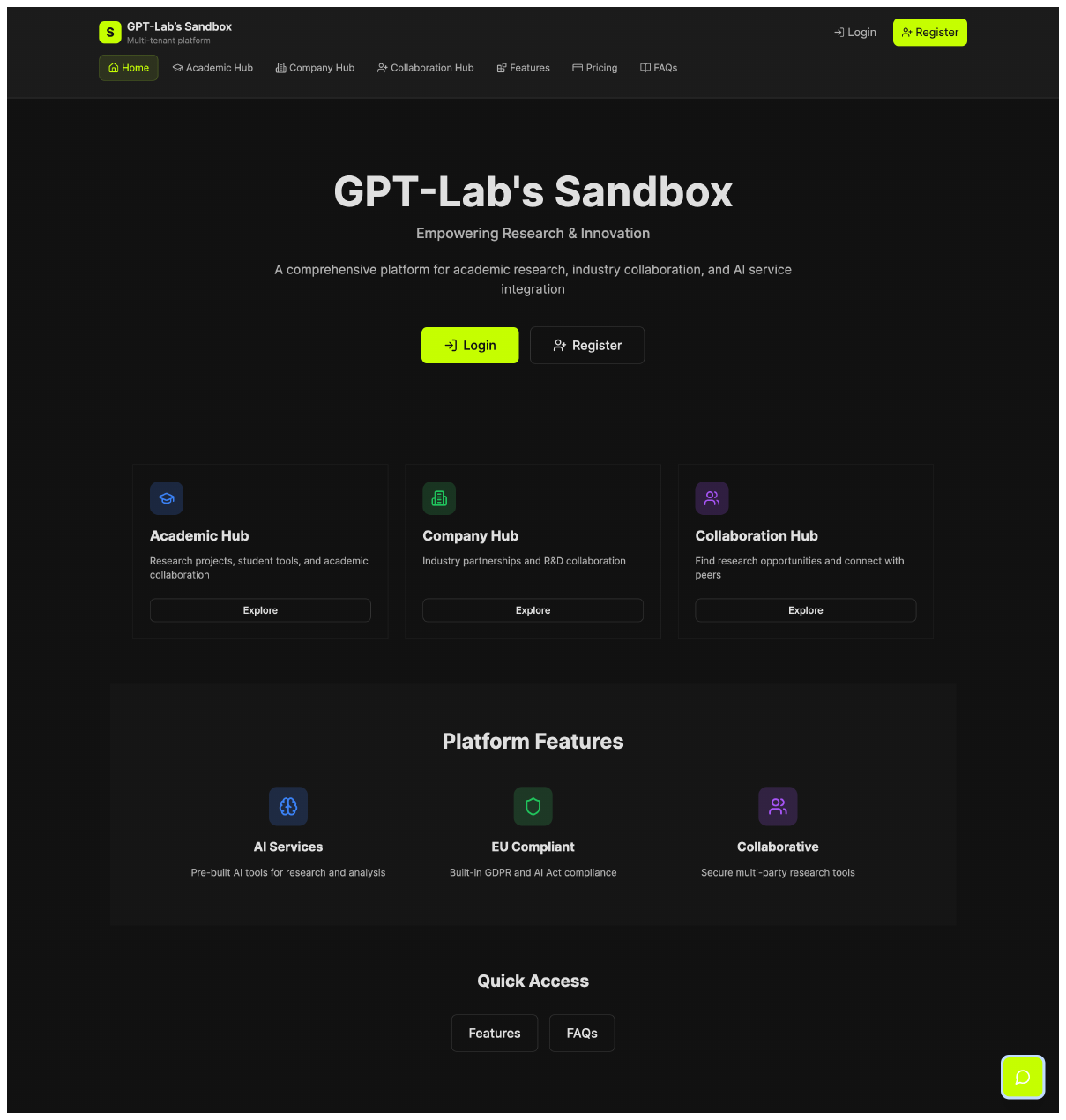}
  \caption[Landing page]{Landing page and secure access gateway of the Sandbox MVP.}
  \label{fig:landingpage}
\end{figure}

\subsection{Academic and Research Hub}

The Academic Hub consolidates institutional projects, datasets, and AI service connections, providing a clear overview of academic participation.

\begin{figure}[H]
  \centering
  \includegraphics[width=0.95\textwidth]{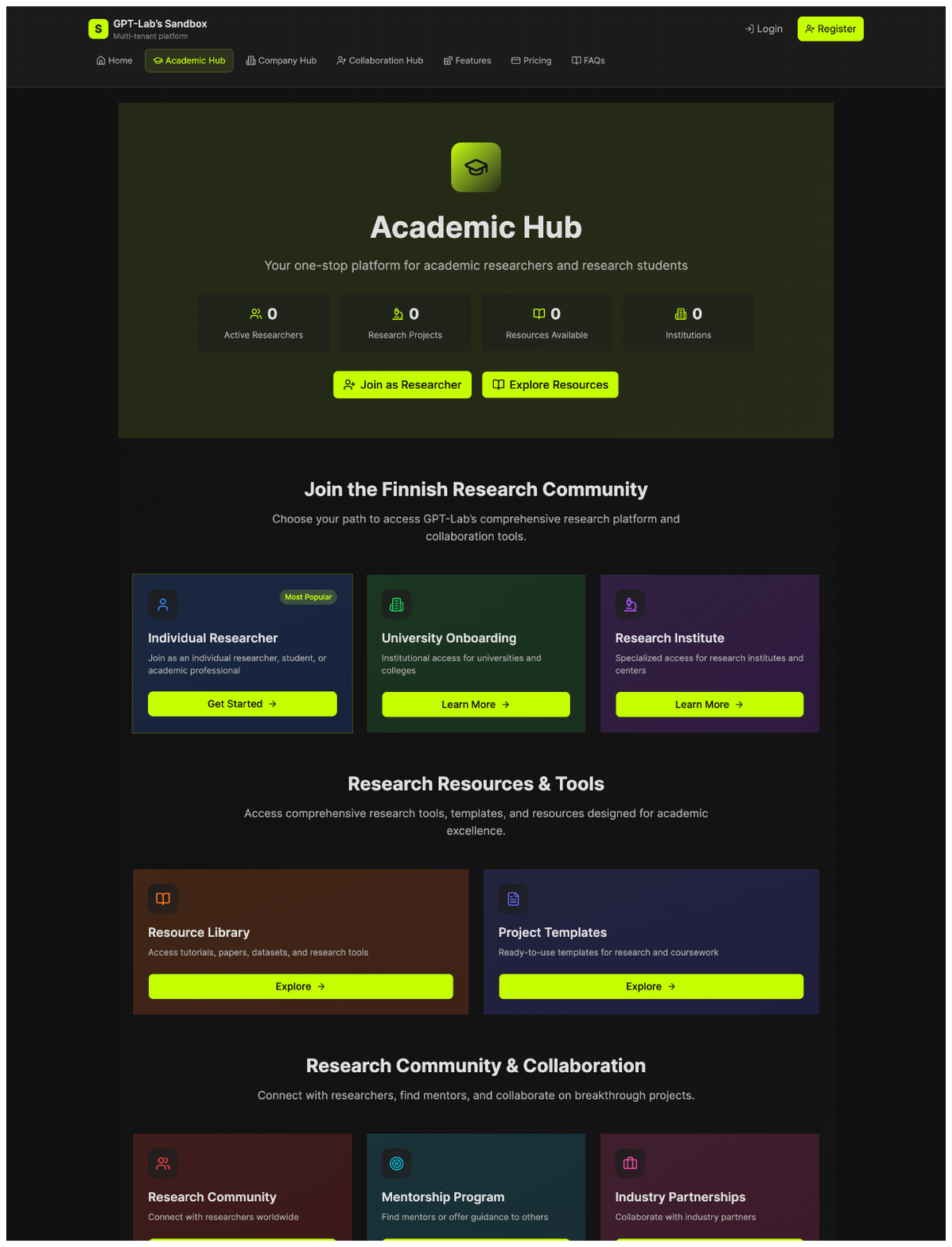}
  \caption[Academic Hub]{Academic Hub view showing institutional projects and datasets}
  \label{fig:academichub}
\end{figure}

\subsection{Company Hub and Collaboration Environment}

The Sandbox supports structured onboarding of companies and industry partners. Each organisation manages its members and projects independently, with isolation and secure data-sharing zones.

\begin{figure}[H]
  \centering
  \includegraphics[width=0.95\textwidth]{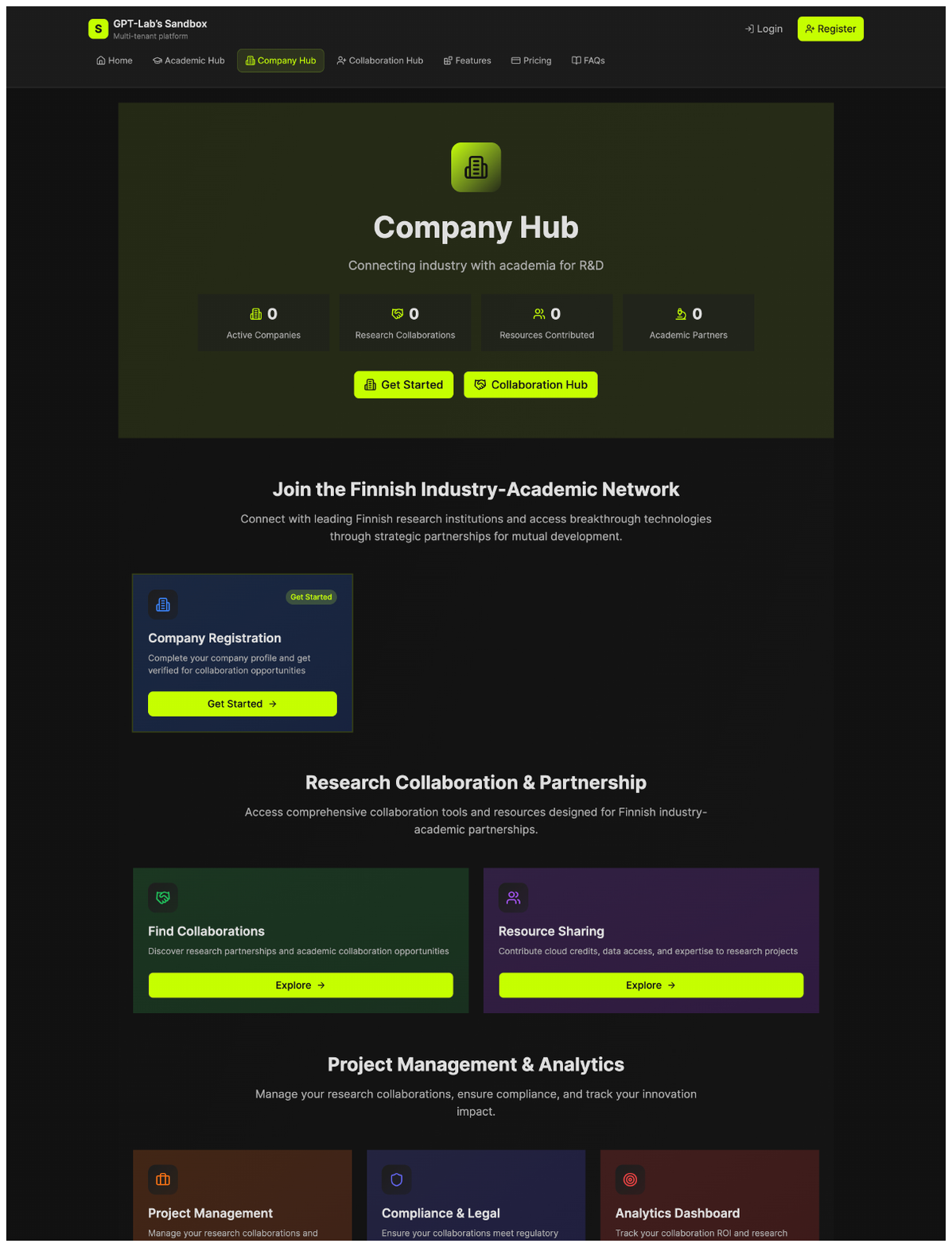}
  \caption[Company Hub]{Company hub displaying industrial participants and collaboration interfaces.}
  \label{fig:companieshub}
\end{figure}

\subsection{System Dashboard and User Management}

The unified dashboard provides administrators with visibility into platform usage and user activities. It integrates account management, approvals, and security role definitions for academic and corporate users.

\begin{figure}[H]
  \centering
  \includegraphics[width=0.95\textwidth]{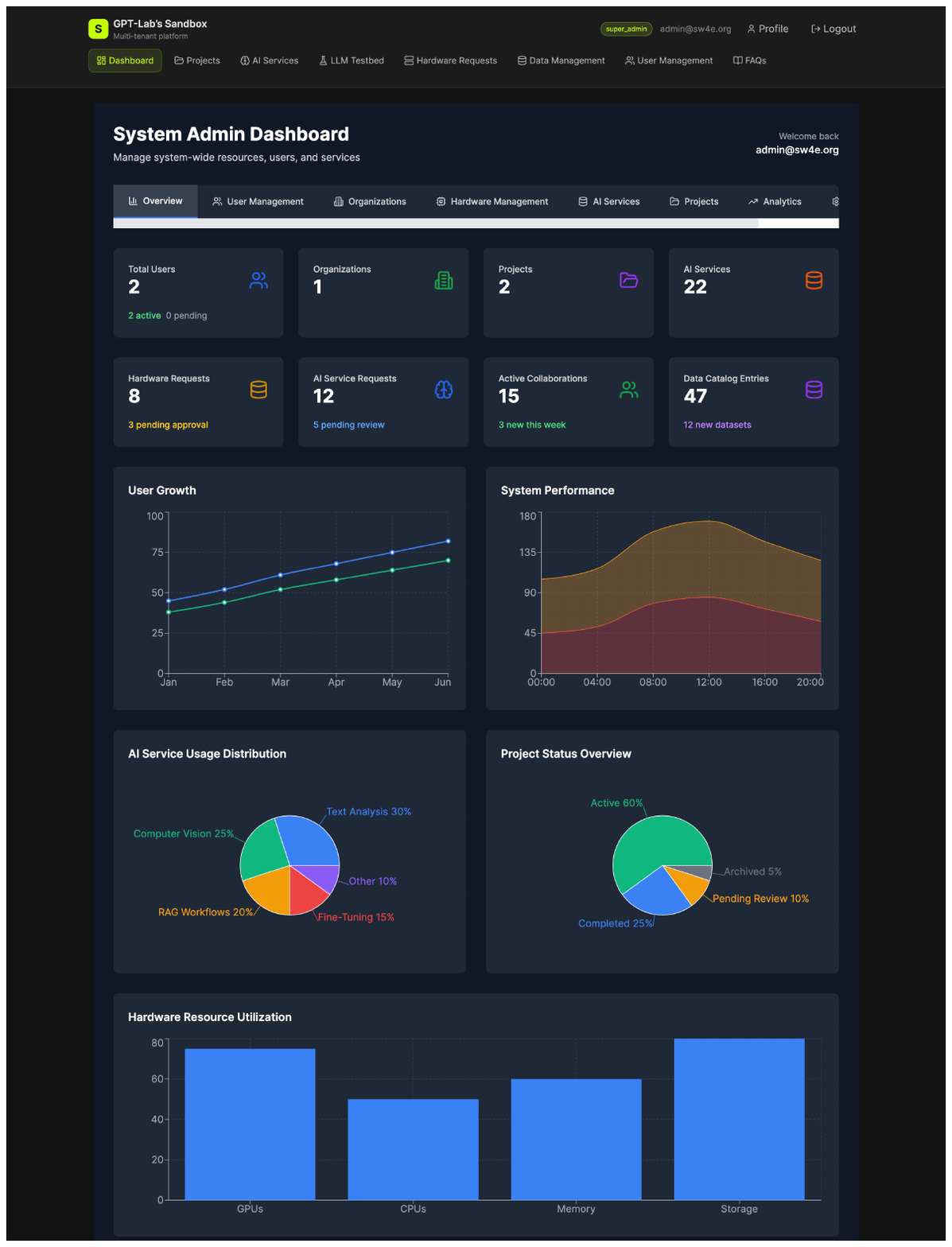}
  \caption[System dashboard]{System dashboard providing visibility across organisational and project layers.}
  \label{fig:dashboard}
\end{figure}


\subsection{Hardware and Resource Management}

Administrators and project leads allocate computational resources through the hardware-management interface, which supports quotas, resource pools, and real-time utilisation monitoring.

\begin{figure}[H]
  \centering
  \includegraphics[width=0.8\textwidth]{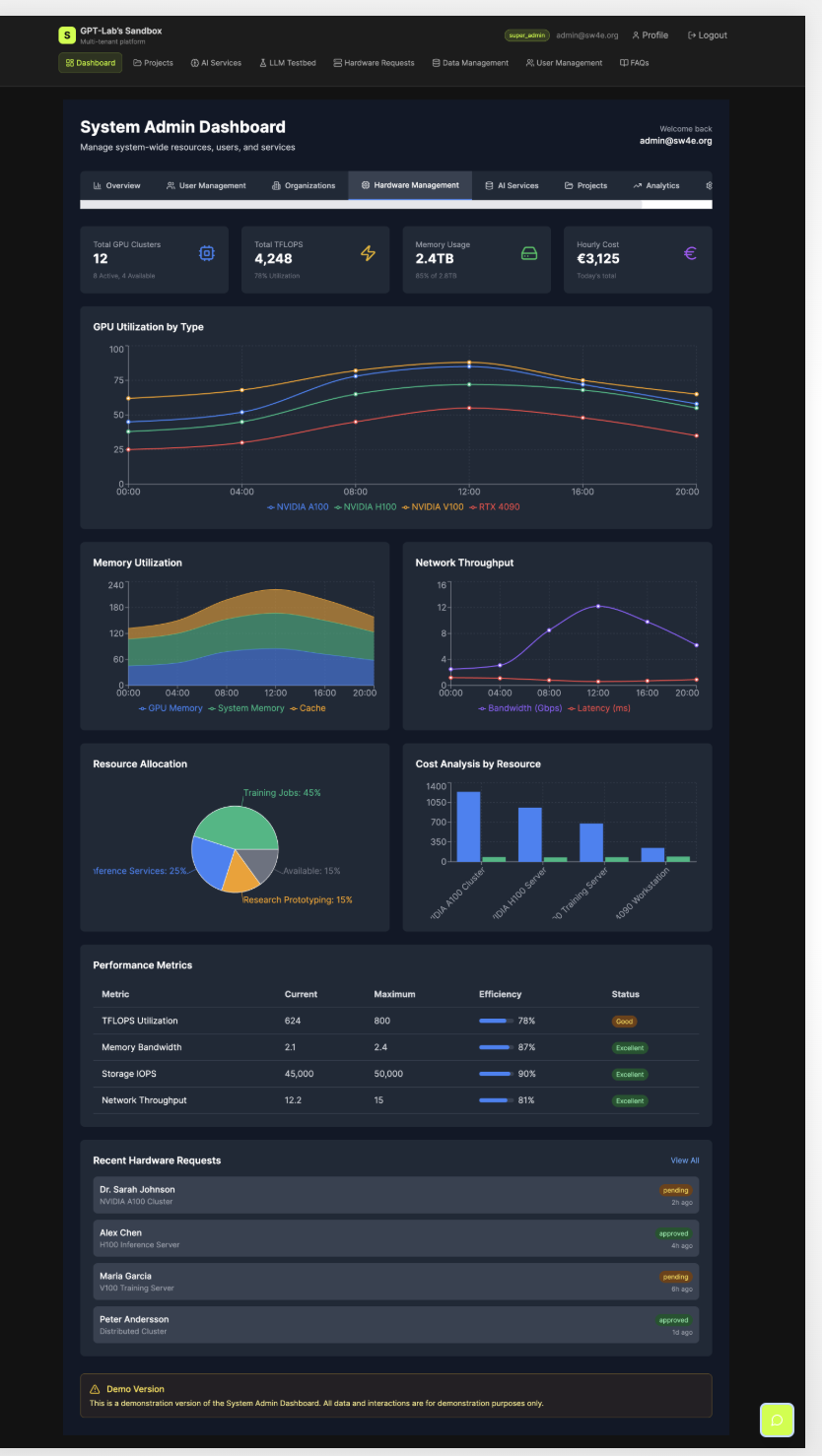}
  \caption[Hardware Management Interface]{Hardware-management interface for GPU/CPU allocation and monitoring.}
  \label{fig:hardware}
\end{figure}

\subsection{AI Services and Model Evaluation}

The AI Services module provides tools for deploying, testing, and benchmarking language models under the Sandbox’s compliance and governance framework.

\begin{figure}[H]
  \centering
  \includegraphics[width=0.95\textwidth]{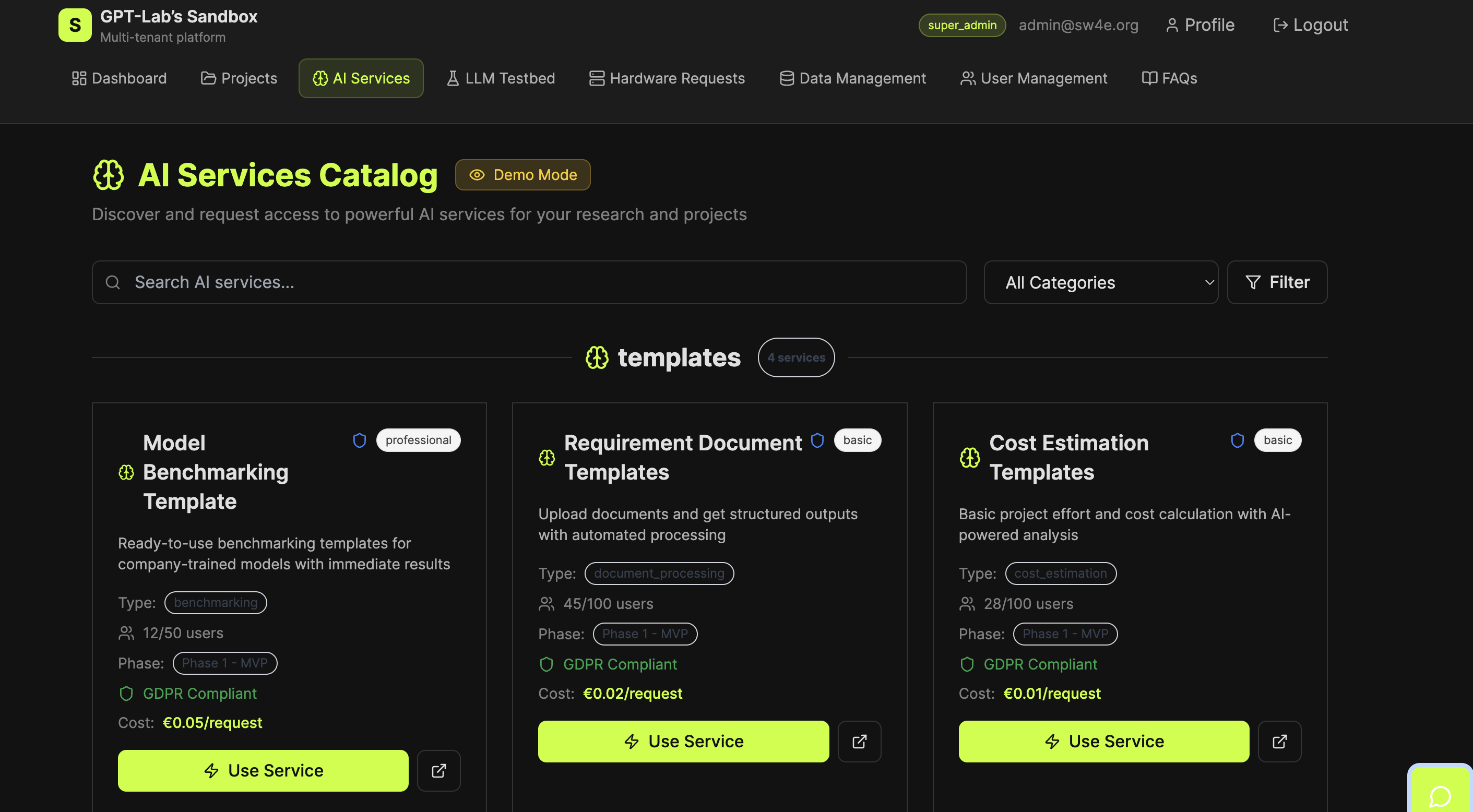}
  \caption[AI Services module]{AI Services module listing available AI models and workflows.}
  \label{fig:aiservices}
\end{figure}

\subsection{LLM Comparison and Benchmarking (One example)}

The Sandbox includes a comparison service that allows users to evaluate multiple LLMs, including open and proprietary options, on shared tasks. 
This module supports evidence-based model selection by comparing response accuracy, latency, and cost-efficiency.

\begin{figure}[H]
  \centering
  \includegraphics[width=0.95\textwidth]{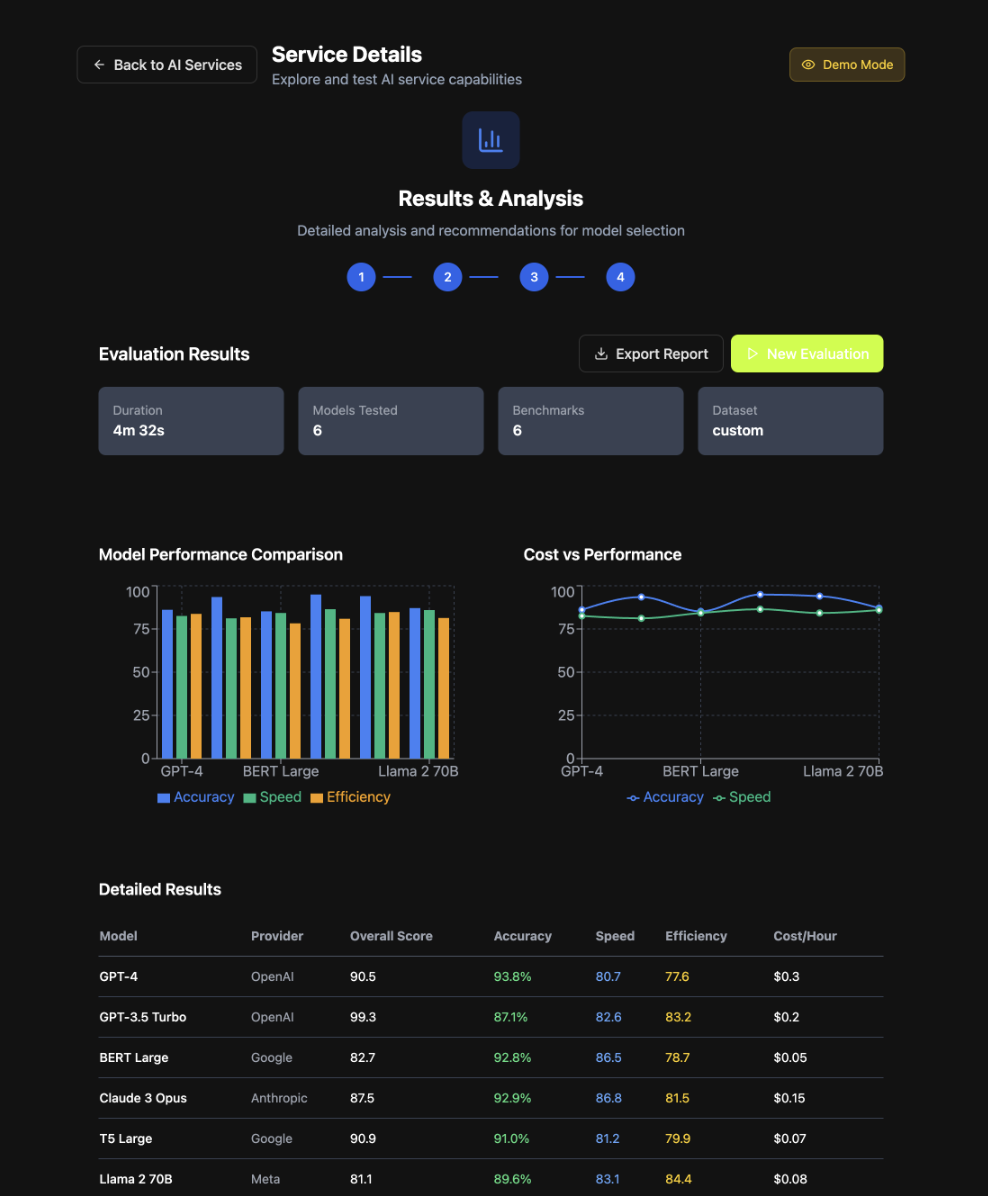}
  \caption[Model benchmarking]{Model benchmarking and comparative evaluation of multiple LLMs within the Sandbox.}
  \label{fig:llmcompare}
\end{figure}

\section{Repository and Documentation Access}

The implementation is available in the GPT Lab GitHub repository, which serves as the source for code, container definitions, and automation scripts:
\begin{quote}
\url{https://github.com/GPT-Laboratory/SW4E-s-Sandbox}
\end{quote}

Comprehensive documentation, including deployment instructions for CSC Rahti, OpenShift, and on-premises environments, is provided in the project Wiki. 
The documentation covers:

\begin{itemize}
  \item Step-by-step deployment and configuration
  \item Integration with institutional identity providers (where applicable)
  \item Role and permission mapping consistent with the governance model
  \item Monitoring, auditing, and security best practices
\end{itemize}

\section{Summary}

The deployed MVP demonstrates the technical feasibility of the AI Sandbox and the practicality of its governance-driven design. 
Its modular, containerised architecture supports extension to additional AI services, while embedded security and audit mechanisms support GDPR compliance and EU AI Act readiness. 
Through its open-source release and documentation in the GPT Lab repository, the Sandbox provides a reusable foundation for organisations seeking to develop, evaluate, and govern AI systems responsibly within shared research and innovation infrastructures.

\chapter{Frequently Asked Questions}
\label{sec:faq}

\noindent
As the Sandbox moves toward real-world deployment, a number of recurring questions arise from companies, data controllers, and research partners regarding compliance, intellectual property, data protection, and accountability. 
This chapter summarises the most frequent concerns and provides concise answers that translate the governance and technical safeguards into practical, real-world assurances.

\begin{enumerate}[leftmargin=*, label=\textbf{Q\arabic*:}]

\item \textbf{How will sensitive business data be protected once uploaded into the sandbox?}\\
All uploaded data is protected under GDPR Article~32 (Security of Processing) and the Finnish Data Protection Act~(1050/2018). 
Data is encrypted in transit (TLS~1.2+) and at rest (AES-256), stored in EU/EEA data centres, and isolated per tenant. 
We apply data-protection-by-design (GDPR~Art.~25) and align our security controls with ISO/IEC~27001-aligned practices and zero-trust principles to ensure end-to-end protection.

\item \textbf{Will encryption and access controls be strong enough to prevent leaks between collaborators?}\\
Yes. The Sandbox enforces strong role-based access control (RBAC), multi-factor authentication (MFA), and strict tenant isolation under GDPR~Art.~32(1) and common NIS2-aligned cybersecurity practices. 
Encryption key management can support customer-managed keys where required by the deployment environment and operating model, ensuring that only authorised parties can decrypt protected data.

\item \textbf{Where will the data be physically stored, and how will GDPR or sector-specific compliance be ensured?}\\
Data residency is enforced within EU/EEA regions (primarily Finland) to comply with GDPR~Chapter~V on international transfers. 
No personal or regulated data will be transferred outside the EEA without a valid transfer mechanism such as an adequacy decision or Standard Contractual Clauses (SCCs). 
Sector-specific rules (e.g., healthcare, finance) are implemented through isolated workspaces, stricter access controls, and documented Data Protection Impact Assessments (DPIAs) under GDPR~Art.~35 where required.

\item \textbf{Who will own the outputs generated by the LLM when multiple organisations contribute data or prompts?}\\
Each participant retains ownership of its own inputs (data, prompts, models). 
Outputs and co-created artefacts follow the project or collaboration agreement, in line with the Finnish Copyright Act~(404/1961) and the EU Trade Secrets Directive~2016/943. 
Unless otherwise agreed in writing, jointly created results will be jointly owned.

\item \textbf{How will proprietary know-how be protected during collaboration?}\\
Proprietary information is safeguarded through tenant and project isolation, confidentiality terms in collaboration agreements, and data-protection-by-design (GDPR~Art.~25). 
Each project operates in an isolated workspace, and cross-project data mixing is not permitted unless explicitly approved and contractually defined. 
Logging and provenance tracking support controlled use and help prevent accidental disclosure of trade secrets.

\item \textbf{How will derivative models or works be handled?}\\
Derivative models and works are governed by explicit licence terms defined in the relevant project or collaboration agreement. 
The Sandbox operator does not claim ownership of participant outputs and does not reuse them outside the agreed scope. 
Practices are aligned with Finnish intellectual property law and applicable EU copyright principles.

\item \textbf{If the LLM produces incorrect, biased, or harmful outputs, who will be responsible?}\\
Responsibility is defined through the applicable controller/processor roles under GDPR and through project-specific contracts. 
The Sandbox maintains traceability through version control, logging, and audit records, supporting accountability and the documentation and oversight expectations associated with EU AI Act risk management, transparency, and human oversight.

\item \textbf{Will companies be able to trace which prompts, datasets, or users produced a specific output?}\\
Yes. Model runs can be logged with user identity, timestamp, dataset reference (for example hashes or identifiers), and model/version metadata. 
These records can be exported for independent audit or regulator review, supporting traceability and accountability.

\item \textbf{How will you ensure explainability and trust in model outputs?}\\
The Sandbox supports documentation and interpretability measures, including dashboards where appropriate. 
Each output can be linked to the model version, dataset reference, and execution context. 
Human-in-the-loop review checkpoints can be applied for sensitive use cases to strengthen transparency and align with AI Act expectations on user information and human oversight.

\item \textbf{How will secure access be managed when companies use different cloud or IT systems?}\\
The Sandbox can support federated identity management via OpenID Connect or SAML~2.0, combined with MFA. 
Access rights are managed through role-based and attribute-based controls, consistent with GDPR~Art.~32(1)(b) and widely adopted ENISA-aligned best practices for cross-organisational security.

\item \textbf{Will collaborators be able to access each other’s internal systems accidentally?}\\
No. The platform is designed to enforce project and tenant boundaries through network segmentation, strict access controls, and API-layer authorisation. 
By design, data and services do not flow between tenants or organisations unless explicitly configured and approved.

\item \textbf{How will collaboration remain smooth without over-exposing internal assets?}\\
The Sandbox provides controlled collaboration mechanisms that allow partners to share datasets or results with fine-grained permissions. 
Core systems and private project assets remain isolated, supporting the purpose-limitation principle (GDPR~Art.~5(1)(b)).

\item \textbf{How will you ensure overall compliance with GDPR and the EU AI Act?}\\
Compliance is embedded by design. 
We conduct Data Protection Impact Assessments (DPIAs) for high-risk processing where required (GDPR~Art.~35) and maintain Records of Processing Activities (Art.~30). 
Compliance reviews and audits are carried out under the agreed governance model, supporting continuous alignment with GDPR obligations and EU AI Act readiness.

\item \textbf{Who will be liable if the system is misused or regulatory breaches occur?}\\
Liability is defined through controller/processor arrangements and project-specific contracts. 
Each participant remains responsible for its own data and actions, while the operator remains responsible for agreed operational and security measures within its role. 
Audit records and incident response procedures support accountability and, where applicable, redress.

\item \textbf{How will export control, sector-specific, or antitrust regulations be addressed?}\\
Projects can be screened against relevant export-control rules (including the EU Dual-Use Regulation~2021/821) and sectoral frameworks where applicable. 
Collaborations are structured to avoid inappropriate information sharing that could raise competition-law concerns (TFEU~Articles~101–102). 
Sensitive datasets are processed within controlled EU/EEA environments, with logging and access controls applied throughout.

\item \textbf{Will audit trails and contracts be maintained for legal defensibility?}\\
Yes. Audit trails, access records, and relevant project documentation are retained according to the retention policy defined in the governance framework and applicable contractual or regulatory requirements. 
Service Level Agreements (SLAs) and Data Processing Agreements (DPAs), where used, include audit and accountability clauses to support defensibility.

\item \textbf{How will cloud cost and resource usage be controlled?}\\
The Sandbox supports transparent usage monitoring with resource caps, quotas, and dashboards. 
These mechanisms help prevent cost overruns and support sustainable operations, including energy-aware practices where they are measurable and applicable.

\item \textbf{How will data quality and trust in partners’ datasets be ensured?}\\
A Data Quality Policy is applied in line with the data governance requirements relevant to the use case, including EU AI Act expectations for data governance for high-risk systems. 
Datasets undergo validation for accuracy, completeness, and lawful provenance before being used for training or analysis, and the outcomes are recorded in dataset metadata where applicable.

\end{enumerate}

\vspace{1em}
\noindent
The questions and answers presented above reflect the Sandbox’s guiding principles of openness, neutrality, accountability, and trust. 
Together they demonstrate how governance, security, and compliance are implemented not merely as policies but as integral technical and organisational features of the platform. 
The following chapter summarises how these principles converge into a sustainable, secure, and transparent model for responsible AI collaboration.

\chapter{Conclusions}

Building upon this foundation, the Sandbox was developed through a structured process that combined stakeholder engagement, architectural design, and iterative validation. Each phase, from requirements analysis to technical implementation and governance formulation, was aligned with the principles of security, transparency, and collaboration. The following development milestones demonstrate how these principles were translated into a functioning, compliant, and extensible platform.

The Sandbox requirements were identified, analysed, and prioritised through consultations with key stakeholders to ensure that data sensitivity, localisation, and regulatory compliance needs were addressed from the outset. This collaborative process brought together technical experts, researchers, and institutional representatives to define functional and non-functional requirements reflecting both operational and ethical considerations. As a result, the Sandbox evolved as a coherent response to the diverse needs of academia, industry, and the public sector, balancing flexibility with accountability and ensuring that user expectations are met within a secure and compliant environment.

The technical architecture of the Sandbox was developed as a modular, container-based system designed for scalability, interoperability, and controlled experimentation. It follows a microservices model with defined layers for orchestration, data management, model services, and security. The technology stack integrates Docker for containerisation, Kubernetes for orchestration, and RESTful APIs for service communication. Core components are implemented using Node.js/Express and Next.js for the prototype control plane and user interface, with SQLite used for prototype persistence and a migration path to PostgreSQL for production-scale deployments. Security-by-design principles are applied throughout, employing encryption, role-based access control, audit logging, and network isolation to support data integrity and system resilience. This combination of technologies and practices enables the platform to meet technical performance requirements while supporting the data protection and cybersecurity expectations associated with European regulation.

The governance and policy framework establishes how the platform operates as a trusted, transparent, and accountable environment. It defines operational roles, access permissions, and data-handling procedures that align with the GDPR, the EU AI Act, and relevant international standards. Governance mechanisms clarify responsibilities across participants, ensuring that activities within the Sandbox remain traceable, auditable, and compliant. This approach promotes fairness and transparency while embedding trust in the operating model. By integrating legal and ethical oversight directly into the system design, the Sandbox treats compliance as a continuous and verifiable process rather than a one-time obligation.

The Minimum Viable Product (MVP) demonstrates the technical and organisational feasibility of secure and collaborative AI experimentation. It provides controlled access to resources, enables cooperative workflows among different institutions, and supports the orchestration of AI services, models, and data pipelines within a defined governance structure. The MVP validates the overall system design and confirms that complex experimentation can occur in an environment that supports privacy, accountability, and reproducibility. It serves as both a proof of concept and a foundation for future expansion, establishing the Sandbox as a functional environment for responsible AI research and innovation.

Overall, the Sandbox demonstrates that responsible and innovative AI experimentation can coexist within a single infrastructure when guided by transparent governance, a secure technological foundation, and sustained stakeholder collaboration. Its architecture and operating model provide a stable base for future scaling, connecting research, policy, and practice through a shared framework of trust, transparency, and technically grounded implementation.


\bibliographystyle{plainurl}
\bibliography{References}

\end{document}